\documentclass[aps,nofootinbib,showkeys,noshowpacs,preprintnumbers,amsmath,amssymb]{revtex4}
\pdfoutput=1

\def\be{\begin{equation}}
\def\ee{\end{equation}}
\def\ba{\begin{eqnarray}}
\def\ea{\end{eqnarray}}
\def\bea{\begin{eqnarray}}
\def\eea{\end{eqnarray}}
\def\bes{\begin{subequations}}
\def\ees{\end{subequations}}
\newcommand{\A}{{\mathcal{A}}}
\newcommand{\tA}{{\widetilde {\mathcal{A}}}}

\newcommand{\tal}{{\widetilde \alpha}}

\newcommand{\MSbar}{\overline{\rm MS}}  
\usepackage{graphics}
\usepackage{graphicx}
\usepackage{dcolumn}
\usepackage{bm}
\usepackage{epsfig}
\usepackage{graphicx,color}
\usepackage{multirow}

\begin{document}

\preprint{FTUV-16-0829, IFIC/16-63, USM-TH-347}

\title{Lattice-motivated holomorphic nearly perturbative QCD}

\author{C\'esar Ayala$^1$}
\author{Gorazd Cveti\v{c}$^2$}
\author{Reinhart K\"ogerler$^3$}

\affiliation{$^1$Department of Theoretical Physics and IFIC, University of Valencia and CSIC, E-46100, Valencia, Spain\\ $^2$Department of Physics, Universidad T{\'e}cnica Federico Santa Mar{\'\i}a, Casilla 110-V, Valpara{\'\i}so, Chile\\ $^3$Department of Physics, Universit\"at Bielefeld, 33501 Bielefeld, Germany}

\date{\today}

\begin{abstract}
Newer lattice results indicate that, in the Landau gauge at low spacelike momenta, the gluon propagator and the ghost dressing function are finite nonzero. This leads to a definition of the QCD running coupling, in a specific scheme, that goes to zero at low spacelike momenta. We construct a running coupling which fulfills these conditions, and at the same time reproduces to a high precision the perturbative behavior at high momenta. The coupling is constructed in such a way that it reflects qualitatively correctly the holomorphic (analytic) behavior of spacelike observables in the complex plane of the squared momenta, as dictated by the general principles of Quantum Field Theories. Further, we require the coupling to reproduce correctly the nonstrange semihadronic decay rate of tau lepton which is the best measured low-momentum QCD observable with small higher-twist effects. Subsequent application of the Borel sum rules to the V+A spectral functions of tau lepton decays, as measured by OPAL Collaboration, determines the values of the gluon condensate and of the V+A 6-dimensional condensate, and reproduces the data to a significantly higher precision than the usual $\MSbar$ running coupling.
\end{abstract}

\maketitle

\section{Introduction}
\label{sec:intr}
The search for an effective QCD running coupling ``constant'' $\alpha_s(-q^2)$, whose $q^2$ dependence is not only specified in the spacelike high-momentum region [by the perturbative renormalization group equation (RGE)], but also in the low-momentum regime, has acquired considerable interest during the last two decades. Since the low-momentum behavior can definitely not be understood within the framework of perturbation theory (which would - among other things - lead to unphysical singularities), nonperturbative approaches have to be applied, the most important ones being lattice calculations. A nonperturbative (lattice) QCD coupling can most conveniently be defined via the ghost gluon coupling, namely as a product of the gluon propagator dressing function and the square of the ghost propagator dressing function in the Landau gauge. Consequently, these propagators  have been one of the main focuses of lattice approximations. The corresponding results, however, are yet rather confusing. Earlier lattice calculations of QCD in the Landau gauge indicated that the gluon propagator in the spacelike infrared (IR) region may go to zero for $q^2 \to  0$, and the ghost propagators are infrared enhanced \cite{LattOld1,LattOld2,LattOld3}. These results were in accordance with the scaling solutions of the Dyson-Schwinger equations (DSE) approach \cite{DSEscale}, and also with the functional renormalization group approach (FRG) \cite{FRG} as well as with the Gribov-Zwanziger approach \cite{Gribovscale}. They would lead to a QCD coupling which is nonzero finite in the IR limit of spacelike momenta. However, more recent lattice calculations, based on larger volumes and higher statistics, indicate that the gluon propagator in the IR limit goes to a nonzero finite value \cite{Lattgluonfinite1} and the ghost propagator is not IR enhanced \cite{Lattghost1,LattcoupNf0,LattcoupNf0b,LattcoupNf2,LattcoupNf4} yielding a ghost dressing function which goes to a finite constant in the IR regime. This behavior is consistent with the decoupling solutions of the DSE approach  \cite{DSEdecoup} and with  the modified Gribov-Zwanziger approach \cite{Gribovdecoup} (cf.~also the FRG approach \cite{FRG}). It leads to a QCD running coupling which goes to zero in the IR limit of spacelike momenta. Because of the higher reliability of these new lattice results we will be guided by the IR behavior of their solutions in the following. Furthermore, we will follow the mentioned definition of the lattice running coupling as the product of the Landau gauge dressing functions.

In the present paper we construct with dispersive approach a running QCD coupling which, on the one hand, reproduces in the spacelike IR regime the main features of the mentioned lattice coupling, and, on the other hand, shows in the  spacelike ultraviolet (UV) regime, to a high precision, the behavior as indicated by perturbative QCD (pQCD). This means that the resulting coupling represents, in a sense, an analytic continuation of pQCD from UV to IR, thereby avoiding unphysical (Landau) singularities, but capturing the major qualitative features of the lattice coupling, and it is in general expected to differ from the latter by nonperturbative terms (corrections). In addition, we request that such a (``nearly perturbative'') coupling reflect the holomorphic (analytic) behavior of the spacelike QCD observables as dictated by the general principles of Quantum Field Theories \cite{BS,Oehme}, since such a coupling is supposed to be used in the evaluation of (the leading-twist part of) such quantities. Therefore, the constructed coupling $\A(Q^2)$, being a function of the squared momentum transfers $q^2 = (q^0)^2 - (q^j)^2 \equiv -Q^2$, is: (I) an analog of the purely perturbative coupling $a(Q^2) \equiv \alpha_s(Q^2)/\pi$ in the same lattice renormalization scheme  (minimal momentum subtraction (MiniMOM) scheme \cite{MiniMOM}), and, (II) in contrast to $a(Q^2)$, the coupling  $\A(Q^2)$ is a holomorphic (analytic) function of $Q^2$ in the generalized spacelike part of the complex $Q^2$-plane, i.e., for $Q^2 \in \mathbb{C} \backslash (-\infty, -M_{\rm thr}^2]$, where $M_{\rm thr} \sim 0.1$ GeV is a threshold mass. There is yet another, phenomenological, condition to be fulfilled by such a coupling: it should reproduce the measured value $r_{\tau} \approx 0.20$, where $r_{\tau}$ is the QCD massless canonical part [$r_{\tau} = a + {\cal O}(a^2)$] of the $\tau$ lepton strangeless total ($V+A$) semihadronic decay ratio; this quantity is at present the principal low-momentum QCD quantity which has been measured to a high precision  (better than $\pm 0.01$) and, simultaneously, is known to have small higher-twist contributions. 

  In Sec.~\ref{sec:latt} we explain the definition of the nonperturbative (lattice) QCD running coupling based on the gluon-ghost-ghost interaction, present the related available lattice results for such a coupling, and describe the (qualitative) features that these results impose on the sought for coupling in the IR spacelike regime. In Sec.~\ref{sec:constr} we construct the coupling, in the MiniMOM scheme, by ensuring the holomorphic behavior, the correct behavior in the UV as well as in the IR regime, and the reproduction of the correct value of the $\tau$ decay ratio $r_{\tau}$. In Sec.~\ref{sec:SR} we briefly describe the application of the obtained $\A$-coupling coupling theory ($\A$QCD) to the Borel sum rules of the ($V+A$) $\tau$-decay spectral functions and present some results, using the Operator Product Expansion (OPE) for the quark current correlator up to the dimension $D=6$ terms. In Sec.~\ref{sec:pred} we present predictions of the $\A$QCD+OPE approach for some QCD low-energy observables, and compare them with those of the usual $\MSbar$ pQCD+OPE approach and with the experimental results. In Sec.~\ref{sec:summ} we summarize the results obtained in this work. The Mathematica scripts for the calculation of the coupling $\A(Q^2)$ and its higher-order analogs is available freely \cite{4danQCDcoupl}.

\section{Nonperturbative (lattice) coupling}
\label{sec:latt}

\begin{figure}[htb] 
\centering\includegraphics[width=100mm]{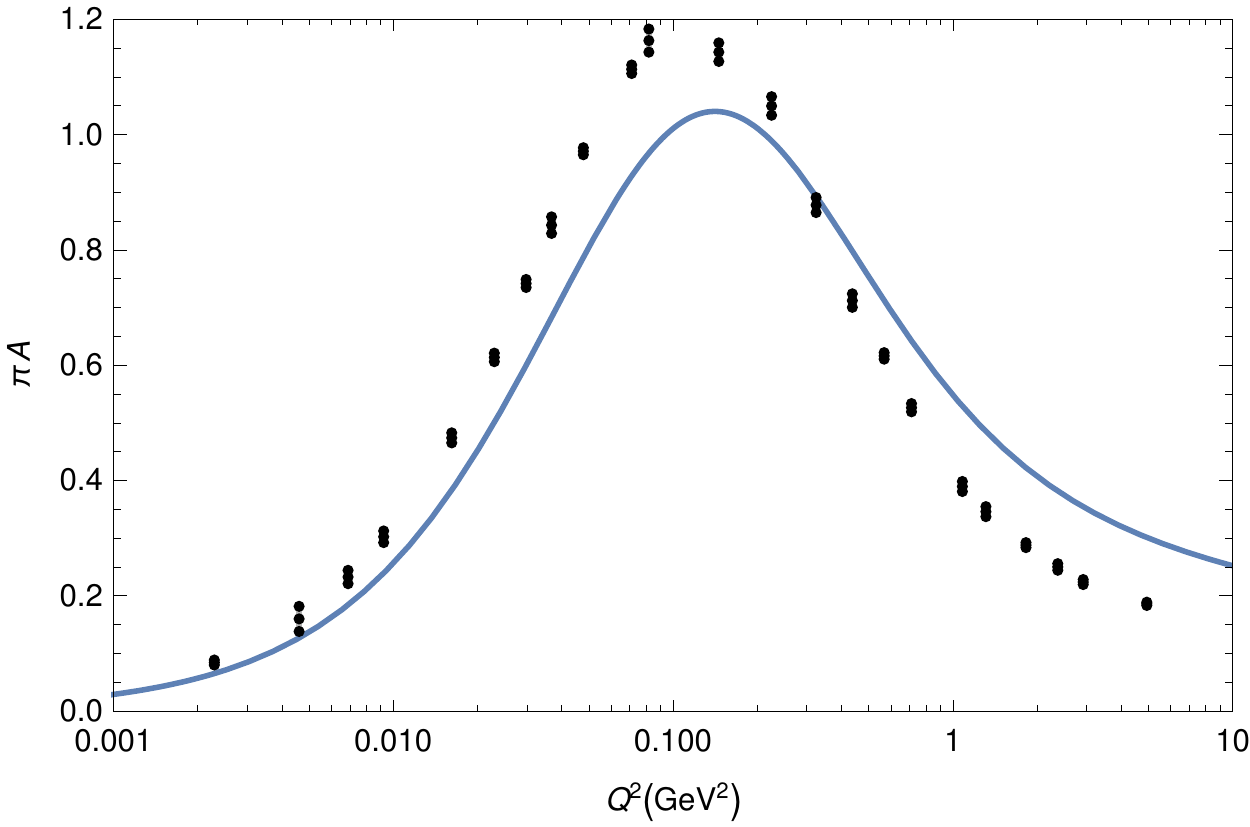}
 \caption{\footnotesize  The nonperturbative (lattice) values $\pi \A_{\rm latt.}(Q^2)$ at low $Q^2$, as obtained in Ref.~\cite{LattcoupNf0}: the (triple) points, which include also calculational uncertainties. The rescaling is performed for the squared momenta, from the MiniMOM (MM) scheme scale to the usual $\MSbar$-like scale: $Q^2 = Q^2_{\rm latt.} ({\Lambda}_{\MSbar}/{\Lambda_{\rm MM}})^2 \approx Q^2_{\rm latt.}/1.9^2$ at $N_f=0$. The continuous curve is our holomorphic coupling (see Sec.~\ref{sec:constr}).}
\label{FigAlatt}
 \end{figure}

In the Landau gauge, the gluon and ghost propagators, in the Wick-rotated formulation ($k^2 \mapsto - K^2$, where $k^2$ is in Minkowski metric, and $K^2$ in Euclidean) and in the theory with UV cutoff $\Lambda$, are 
\bes
\label{Ps}
\bea
\left({\cal P}_{\rm gl}^{(\Lambda)} \right)^{ab}_{\mu \nu}(K) & = & \delta^{ab}
Z^{(\Lambda)}_{\rm gl}(K^2) \frac{1}{K^2} \left( \delta_{\mu \nu} -
\frac{K_{\mu} K_{\nu}}{K^2} \right) \ ,
\label{Pgl} 
\\
{\cal P}_{\rm gh}^{(\Lambda)}(K) & = & 
- Z^{(\Lambda)}_{\rm gh}(K^2) \frac{1}{K^2} \ ,
\label{Pgh} 
\eea
\ees
where the superscript $(\Lambda)$ indicates here that the theory is regularized by UV momentum cutoff $\Lambda^2$. The dressing functions $Z^{(\Lambda)}_{\rm gl}(K^2)$ and $Z^{(\Lambda)}_{\rm gh}(K^2)$ appear in the nonperturbative (lattice) definition of the running coupling
\bea
\A_{\rm latt.}(Q^2) & = & \A_{\rm latt.}(\Lambda^2) Z^{(\Lambda)}_{\rm gl}(Q^2) \left(Z^{(\Lambda)}_{\rm gh}(Q^2) \right)^2 \ ,
\label{Alatt}
\eea
This identity is based on the fact that the gluon-ghost vertex renormalization parameter ${\widetilde Z}_1^{(\Lambda)}(Q^2)$ in the Landau gauge is equal to one, at all energies,  i.e., nonrenormalization of the gluon-ghost-ghost vertex in the Landau gauge,
Ref.~\cite{Taylor} (cf.~also Refs.~\cite{DSEscale}).\footnote{In general we would have: 
$a(Q^2) \propto Z_{\rm gl}^{(\Lambda)}(Q^2) Z_{\rm gh}^{(\Lambda)}(Q^2)^2/{\widetilde Z}_1^{(\Lambda)}(Q^2)^2$.}  Since the gluon propagator has a finite nonzero limit at $K^2 \downarrow 0$ \cite{Lattgluonfinite1}, we have $Z^{(\Lambda)}_{\rm gl}(Q^2) \sim Q^2$ when $Q^2 \downarrow 0$.
There exist now large volume and high statistics lattice calculations \cite{LattcoupNf0,LattcoupNf0b} for the gluon and ghost propagators in the Landau gauge in the quenched ($N_f=0$) approximation, which show that $Z^{(\Lambda)}_{\rm gh}(Q^2) \to {\rm const}$ when $Q^2 \downarrow 0$, i.e., that ghost propagator is not IR enhanced, in contrast to the results of earlier lattice calculations. As a consequence, the nonperturbative (lattice) coupling (\ref{Alatt}) goes to zero, $\A_{\rm latt.}(Q^2) \sim Q^2$, when $Q^2 \downarrow 0$. These results are represented as points in Fig.~\ref{FigAlatt}.
Qualitatively similar results are obtained also for unquenched cases, \cite{LattcoupNf2} ($N_f=2$), \cite{LattcoupNf4} ($N_f=4$), although the lattice volume and the statistics are in general smaller there. The renormalization scheme and the scaling in which the lattice calculations are performed is the minimal momentum subtraction (MiniMOM) scheme \cite{MiniMOM} (cf.~Ref.~\cite{KatMol} for a discussion and an application of this scheme).
Since the MiniMOM scheme involves, in addition, a rescaling ($\Lambda_{\rm MM} \approx 1.9 \Lambda_{\MSbar}$ for $N_f=0$ \cite{MiniMOM}), we rescaled the results of Ref.~\cite{LattcoupNf0} back to the usual scale convention ($\Lambda_{\MSbar}$) in Fig.~\ref{FigAlatt}. Very similar results to those of Ref.~\cite{LattcoupNf0} were obtained recently in Ref.~\cite{LattcoupNf0b}, where the authors used different lattice volumes and spacings.
Further, similar results for the QCD coupling [namely, $\A(0)=0$] were obtained also by recent ($N_f=0$) three-gluon vertex lattice calculations \cite{Latt3gluon} and theoretically explained there by solutions of the DSE equations.

In general we expect that the combination $\A_{\rm latt.}(Q^2)$ of dressing functions, Eq.~(\ref{Alatt}), differs from the formal perturbative coupling (in MiniMOM) $a(Q^2)$  at $Q^2 \sim 1 \ {\rm GeV}^2$ by higher dimensional (higher-twist) terms, cf.~\cite{LattcoupNf4}. On the other hand, our goal is to construct a holomorphic nearly perturbative coupling $\A(Q^2)$ which is, in a sense, an analytical continuation of the (MiniMOM) pQCD coupling $a(Q^2)$ into the IR regime. We thus expect
\bea
\A_{\rm latt.}(Q^2) & = & \A(Q^2) + \Delta \A_{\rm NP}(Q^2) 
\label{AlattA}
\eea
for all  $Q^2$,
where $\Delta \A_{\rm NP}(Q^2)$ represents the mentioned nonperturbative (NP, ``higher-twist,'' etc.) corrections. Since $\A_{\rm latt.}(Q^2)$ goes to zero as $Q^2$ when $Q^2 \downarrow 0$, the relation (\ref{AlattA}) implies that the holomorphic coupling $\A(Q^2)$ also has the same behavior in deep IR, 
\be
\A(Q^2) \sim Q^2 \qquad (Q^2 \to 0) \ ,
\label{limA}
\ee
because otherwise we would have a problem of fine-tuning in the relation (\ref{AlattA}). Specifically, if we had $\A(Q^2) \to \A(0) \not= 0$ when $Q^2 \downarrow 0$, then the nonperturbative correction $\Delta \A_{\rm NP}(Q^2)$ would have to tend, when $Q^2 \downarrow 0$, to the nonzero value, $\Delta \A_{\rm NP}(Q^2) \to - \A(0)$ with high precision, this representing a fine-tuning. If we rewrite Eq.~(\ref{AlattA}) in the factorized form $\A_{\rm latt.}=\A(1 + \Delta {\widetilde \A}_{\rm NP})$, where $\Delta {\widetilde \A}_{\rm NP}$ is the relative NP correction (to be compared with OPE higher-twist when $Q^2 > \Lambda^2_{\rm QCD}$), then the fine-tuning means that $\Delta {\widetilde \A}_{\rm NP} \to -1$ to high precision when $Q^2 \to 0$. No fine-tuning means generally that $\Delta {\widetilde \A}_{\rm NP} \not\to  -1$ when $Q^2 \to 0$.
Therefore, we regard the IR behavior (\ref{limA}) as the main condition coming from lattice calculations that we impose on the holomorphic coupling $\A(Q^2)$ in the deep IR regime. This, in conjunction with the fact that $\A(Q^2)$ at higher positive $Q^2$ ($ > \Lambda^2$) is a monotonically decreasing function [since we will require it to be practically equal to the pQCD coupling $a(Q^2)$ in MiniMOM scheme there], implies that $\A(Q^2)$ has a maximum at $Q^2 \sim \Lambda^2$ ($\sim 0.1 \ {\rm GeV}^2$). Thus, $\A(Q^2)$ should behave qualitatively in a similar way as $\A_{\rm latt.}(Q^2)$ of Fig.~\ref{FigAlatt} in the IR regime.

Generally, one could imagine that the maximum of $\A_{\rm latt}(Q^2)$ and $\A(Q^2)$ is related to the hadronization phenomenon.

\section{Construction of the nearly perturbative holomorphic coupling}
\label{sec:constr}

Various QCD couplings $\A(Q^2)$ which are holomorphic in the complex $Q^2$-plane in the generalized spacelike regime $Q^2 \in \mathbb{C} \backslash (-\infty, -M_{\rm thr}^2]$ have been contructed in the literature, especially since mid-nineties, among them the Analytic Perturbation Theory (APT, a minimal analytic model) \cite{ShS,MS,Sh1Sh2}, its extension to any physical quantity \cite{KS} and to analogs of noninteger powers of the coupling \cite{BMS} (Fractional Analytic Perturbation Theory - FAPT). For reviews of these approaches, we refer to Refs.~\cite{Prosperi,Shirkov,Bakulev,Stefanis}. Some of the applications of these works to low-momentum QCD phenomenology are given in Refs.~\cite{APTappl}.

  Several other models leading to holomorphic couplings have been constructed and applied since then, cf.~Refs.~\cite{Nest2,Webber,CV12,Alekseev,1danQCD,2danQCD,anOPE,anOPE2,Brod,ArbZaits,Boucaud,Shirkovmass,KKS,Luna}, all having finite values of $\A(0)$ (IR ``freezing''), while the construction of Refs.~\cite{Nest1} leads to holomorphic coupling infinite at the origin $\A(0) = \infty$. Mathematical packages for numerical evaluation of various holomorphic (analytic) couplings  and their power analogs are given in Refs.~\cite{BK,ACprogr}, and some of the reviews in Refs.~\cite{GCrev,Brodrev}.
    Most of the constructions involve the use of the Cauchy theorem (dispersive integral approaches) applied to the couplings to ensure that they are holomorphic. Further, related dispersive approaches have been applied also directly to spacelike QCD observables, \cite{MSS1,MSS2,MagrGl,mes2,DeRafael,MagrTau,Nest3a,Nest3b}, for a review cf.~\cite{NestBook}. All these approaches generate in the couplings and/or observables, in addition to the purely perturbative terms, also nonperturbative terms such as power corrections $1/(Q^2)^N \propto \exp[- N B/a(Q^2)]$ [we note that $ \exp(- N B/a)$ has an essential singularity at $a=0$].

On the other hand, there exist also renormalization schemes in pure pQCD\footnote{These are schemes in which the beta function $\beta(a)$ is a function [of pQCD coupling $a(Q^2)$] which is Taylor-expandable around the point $a=0$.} which also result in holomorphic couplings in the regime $Q^2 \in \mathbb{C} \backslash (-\infty, -M_{\rm thr}^2]$  and at the same time reproduce the QCD low-momentum $\tau$-lepton decay phenomenology \cite{anpQCD1a,anpQCD1b,anpQCD2}.

  Most of the holomorphic running couplings $\A(Q^2)$ in the literature have a freezing behavior in the IR regime, i.e., $\A(0)$ is finite positive. However, couplings with the property $\A(0)=0$ have been constructed and physically motivated in the literature \cite{ArbZaits,Boucaud,mes2} independently of the lattice results \cite{LattcoupNf0,LattcoupNf2,LattcoupNf4,Latt3gluon} and of the results of the DSE and Gribov-related aproaches \cite{DSEdecoup,Latt3gluon,Gribovdecoup} which also gave $\A(0)=0$. 

Furthermore, the authors of Ref.~\cite{DSEdecoupFreez} defined the running coupling in the infrared regime in a specific way which results in $\A(0)$ being finite positive even when the Landau gauge gluon and ghost propagators have decoupling solutions. This was done by using a form of the gluon propagator with an effective dynamical gluon mass $M(Q^2)$ in the infrared. In this approach, an effective gluon mass enters in the propagator and in the coupling parameter. On the other hand, in the approach of Eqs.~(\ref{Alatt})-(\ref{limA}), the principal effects of such an effective gluon mass are contained in the coupling. The two definitions of the coupling can possibly be made equivalent in calculations of physical quantities, by using for the gluon propagators the corresponding expressions. In our approach, as a consequence, when the squared momentum $Q^2$ decreases below the hadronization scale, the residual interaction gradually turns off, cf. Fig.~\ref{FigAlatt} and Eq.~(\ref{limA}). Here we will not follow the line of Ref.~\cite{DSEdecoupFreez}, but rather the line of Eqs.~(\ref{Alatt})-(\ref{limA}).
  
  The basic idea of constructing a coupling $\A(Q^2)$ holomorphic in $Q^2 \in \mathbb{C} \backslash (-\infty, -M_{\rm thr}^2]$ is the following: apply the Cauchy theorem to the integrand $\A(Q'^2)/(Q'^2 - Q^2)$ in the complex $Q'^2$-plane and then use asymptotic freedom [$\A(Q'^2) \to 0$ when $Q'^2 \to \infty$], leading 
to the following dispersive integral expression for $\A(Q^2)$:
\bea
\A(Q^2) & = & \frac{1}{\pi} \int_{M_{\rm thr}^2- \eta}^{+\infty} d \sigma \frac{\rho_{\A}(\sigma)}{(\sigma + Q^2)} \ , \qquad (\eta \to +0)
\label{Adisp}
\eea
where $\rho_{\A}(\sigma) \equiv {\rm Im} \A(-\sigma - i \epsilon)$ is the discontinuity function of $\A(Q^2)$ along its cut. In general, $\rho_{\A}(\sigma)$ is set to be equal to the perturbative  discontinuity function at large $\sigma > \Lambda^2$ (where $\Lambda^2 \sim 0.1 \ {\rm GeV}^2$),\footnote{
We note that the perturbative discontinuity function $\rho_a(\sigma) \equiv {\rm Im} a(-\sigma - i \epsilon)$ is obtained from the underlying pQCD coupling $a(Q^2)$ which is in a chosen renormalization scheme which defines also the renormalization scheme of $\A(Q^2)$. We refer to $a(Q^2)$ as the ``underlying'' pQCD coupling.} and at low positive $\sigma$ the (a priori unknown) behavior of $\rho_{\A}(\sigma)$ is parametrized as a combination of delta functions. This leads via Eq.~(\ref{Adisp}) to a holomorphic function which tends to the underlying pQCD coupling $a(Q^2)$ at large $|Q^2|$. The idea of parametrizing the low-$\sigma$ regime of $\rho_{\A}$ as a linear combination of delta function was implemented in the works \cite{1danQCD,2danQCD}. It is motivated by the fact that a linear combination of delta functions $\pi {\cal F}_j \delta(\sigma - M_j^2)$ in $\rho_{\A}(\sigma)$ corresponds to a linear combination of simple fractions ${\cal F}_j/(Q^2+M_j^2)$ in $\A(Q^2)$, which represents an off-diagonal Pad\'e of the type $[N-1/N](Q^2)$ which usually can approximate any holomorphic function $\Delta \A(Q^2)$ to an increasing precision when the number ($N$) of deltas increases. This convergence is even guaranteed if the holomorphic function $\Delta \A(Q^2)$ is a Stieltjes function [i.e., with positive definite discontinuity function $\rho(\sigma)$], \cite{Peris,Pade}. However, since we will require that, in the deep IR regime, our (nearly perturbative) holomorphic coupling $\A(Q^2)$ reproduces qualitatively the main features of the lattice coupling behavior, Fig.~\ref{FigAlatt}, i.e., a nonmonotonic behavior, it turns out to be efficient to assume that our low-$\sigma$ parametrization of $\rho_{\A}(\sigma)$ contains, instead of a linear combination of three or four delta functions, basically delta functions around two low-$\sigma$ points, at one of them such combinations of deltas which simulate the first and the second derivative:\footnote{
We tried to fulfill all the imposed conditions first with four delta functions at four different (and variable) places in low-$\sigma$ regime, but it turned out that the parametrization Eq.~(\ref{rhoA}) was the most efficient one to satisfy all those conditions.}
\bes
\label{rhoA}
\bea
\rho_{\A}(\sigma) & = & \Delta \rho_{\A}(\sigma) + \Theta(\sigma - M_0^2) \rho_a(\sigma) \ ,
\label{rhoAa}
\\
 \frac{1}{\pi} \Delta \rho_{\A}(\sigma) & = &
{\cal F}_1 \delta(\sigma - M_1^2) + {\cal F}_2 \delta(\sigma - M_2^2) 
+ {\cal F}_1^{(1)}  \delta'(\sigma - M_1^2) + {\cal F}_1^{(2)}  \delta''(\sigma - M_1^2)
\label{rhoAb}
\eea
\ees
Here, $\Theta$ is the Heaviside step function and $M_0^2$ is the ``pQCD-onset'' scale, and $\rho_a(\sigma) = {\rm Im} a(-\sigma - i \epsilon)$ is the discontinuity function of the underlying pQCD coupling $a(Q^2) = \alpha_s(Q^2)/\pi$, in the MiniMOM renormalization scheme \cite{MiniMOM} but with the usual ($\MSbar$-like) scaling ${\Lambda}_{\MSbar}$ and the number of active quark flavors considered to be $N_f=3$.\footnote{Since the considered low-momentum physics will be $\tau$-decay physics, and since the current masses of the first three quarks are all $\alt 0.1$ GeV, we will take $N_f=3$ at all the considered momenta.} 
Application of the dispersion integral (\ref{Adisp}) together with the discontinuity function (\ref{rhoA}) yields the following form for the (holomorphic) coupling $\A(Q^2)$:
\bes
\label{AQ2}
\bea
\A(Q^2) & = &  \Delta \A(Q^2) + \frac{1}{\pi} \int_{M_0^2}^{\infty} d \sigma \frac{ \rho_a(\sigma) }{(Q^2 + \sigma)} \ ,
\label{AQ2a}
\\
\Delta \A(Q^2) & = &
\sum_{j=1}^2 \frac{{\cal F}_j}{(Q^2 + M_j^2)} + \frac{{\cal F}_1^{(1)} }{(Q^2 + M_1^2)^2} + \frac{2 {\cal F}_1^{(2)}}{(Q^2 + M_1^2)^3} \ .
\label{AQ2b}
\eea
\ees
We note that $\A(Q^2)$ is not a Stieltjes function, i.e., $\rho_{\A}(\sigma)$ is not positive definite [$\rho_a(\sigma)$ is]. This means that at least one of the parameters ${\cal F}_j$, ${\cal F}_1^{(1)}$, ${\cal F}_1^{(2)}$ must be negative. This is so because otherwise $\A(Q^2)$ would be monotonically decreasing function for all positive $Q^2$ [$\partial \A(Q^2)/\partial Q^2 < 0$], which would contradict lattice results Fig.~\ref{FigAlatt}.

The MiniMOM renormalization scheme has been determined at the 4-loop level in Ref.~\cite{MiniMOM}, in its mass independent variant, with the scheme parameters $c_j \equiv \beta_j/\beta_0$ at $N_f=3$ acquiring the values $c_2=9.3$ and $c_3=71.45$. The parameters $c_j$ enter in the renormalization group equation
\be
\frac{d a(Q^2)}{d \ln Q^2} = - \beta_0 a(Q^2)^2 \left[ 1 + c_1 a(Q^2) + c_2 a(Q^2)^2 + c_3 a(Q^2)^3 + \ldots \right] \ ,
\label{RGE}
\ee
where the coefficients $\beta_0 =(1/4)(11 - 2 N_f/3)$ and 
$c_1 = \beta_1/\beta_0 =(1/4) (102-38 N_f/3)/(11-2 N_f/3)$ are universal in
the mass independent schemes, and the coefficients $c_j$ ($j \geq2$) determine the renormalization scheme \cite{Stevenson}.  However, we will choose a so called Lambert renormalization scheme (Lambert-MiniMOM), i.e., with beta function in the form of a Pad\'e with  $c_2=9.3$, which is equal to MiniMOM only at 3-loop level, and without the (trivial) rescaling $\Lambda_{\MSbar} \mapsto \Lambda_{\rm MM}$. The reason for that is that in such a 3-loop Lambert-MiniMOM scheme, practical evaluation of $\rho_a(\sigma)$ at high $\sigma$ is much more precise and numerically stable (we are using Mathematica software, \cite{Math}), because the RGE in such a scheme has an explicit and simple solution (in terms of Lambert functions). Namely, for this scheme, the RGE is
\bea
\frac{d a(Q^2;c_2)}{d \ln Q^2} & = &
- \beta_0 a^2 \frac{\left[ 1 + (c_1 - (c_2/c_1)) a
\right]}{\left[ 1 - (c_2/c_1) a \right]} \ ,
\label{RGELamb}
\eea
with $c_2=9.3$.
The expansion of the above beta function $\beta(a) =
\partial a/\partial \ln Q^2$ in powers of $a$ is
\be
\beta(a) = 
- \beta_0 a^2 
\left( 1 + c_1 a + c_2 a^2 + \frac{c_2^2}{c_1} a^3 + \ldots \right) \ ,
\label{betapt}
\ee
which implies that in this scheme the four-loop parameter is $c_3=c_2^2/c_1 = 48.7$, which differs somewhat from $c_3=71.45$ of the exact 4-loop MiniMOM scheme.
The explicit solution to RGE (\ref{RGELamb}) is \cite{Gardi:1998qr}\footnote{For explicit solution of RGE beyond three-loops, i.e., not just with given free $c_2$ but also with given free $c_3$ and even $c_4$, see Ref.~\cite{GCIK}; these solutions are more involved and lead to numerical evaluations which require more time.}
 \bea
a(Q^2;c_2) = - \frac{1}{c_1} \frac{1}{\left[
1 - c_2/c_1^2 + W_{\mp 1}(z) \right]} \ ,
\label{aptexact}
\eea
where $Q^2=|Q^2| \exp(i \phi)$; the functions $W_{-1}$ and $W_{+1}$
are the branches of the Lambert function
for $0 \leq \phi < + \pi$ and $- \pi < \phi < 0$, 
respectively, and the variable $z$ is a function of $Q^2$ and
the Lambert scale ${\Lambda}_{\rm L}$
\be
z =  - \frac{1}{c_1 e} 
\left( \frac{|Q^2|}{{\Lambda}_{\rm L}^2} \right)^{-\beta_0/c_1} 
\exp \left( - i {\beta_0}\phi/c_1 \right) \ .
\label{zexpr}
\ee 
In Mathematica \cite{Math}, the functions $W_{\pm 1}(z)$ are implemented with high precision (as  ${\rm ProductLog}[\pm 1,z]$). The formula (\ref{aptexact}) allows us to evaluate efficiently $\rho_a(\sigma) \equiv {\rm Im} a(-\sigma - i \epsilon)$ for all $\sigma$.

A reference value of the Lambert-MiniMOM scheme pQCD coupling $  (Q^2)$ of Eq.~(\ref{aptexact}) is obtained in the following way. The 2014 world central average value $a(M_Z^2,\MSbar) = 0.1185/\pi$ \cite{PDG2014} is used as the starting point, in the considered central case. This value is then RGE-evolved to low energies by 4-loop $\MSbar$ RGE beta function, using at the quark thresholds $Q^2 = (2 {\overline m}_q)^2$ the corresponding 3-loop threhold relations \cite{CKS}. This gives us at $Q^2 = (2 {\overline m}_c)^2$  ($\approx 6.45 \ {\rm GeV}^2$) and $N_f=3$ the value $a(\MSbar) =0.26589/\pi$. The change of the renormalization scheme to the Lambert-MiniMOM (and no rescaling  $\Lambda_{\MSbar} \mapsto \Lambda_{\rm MM}$) is then performed according to relations of Ref.~\cite{Stevenson} [cf.~also App.~A of \cite{BSRGCRK}], giving in this scheme the value $a =0.28043/\pi$ at $Q^2 = (2 {\overline m}_c)^2$ and $N_f=3$, and the Lambert scale value ${\Lambda}_{\rm L}=1.153$ GeV. For more details on this procedure, we refer to \cite{2danQCD,anpQCD2,ACprogr,inprep}. We point out that we construct our coupling $\A(Q^2)$ in (Lambert-)MiniMOM renormalization scheme, and not in the usual $\MSbar$ scheme, in order to make the comparison of $\A(Q^2)$ with $\A_{\rm latt.}(Q^2)$ at low momenta $Q^2 < 1 \ {\rm GeV}^2$, Fig.~\ref{FigAlatt}. Since the latter regime is deep IR and therefore includes nonperturbative contributions, we are not able to make an unambiguous change from MiniMOM to $\MSbar$ in that regime, neither for $\A(Q^2)$ nor for  $\A_{\rm latt.}(Q^2)$.  

Once having the reference value of the Lambert-MiniMOM underlying pQCD coupling (\ref{aptexact}), and thus having $\rho_a(\sigma) \equiv {\rm Im} a(-\sigma - i \epsilon)$, many of the free parameters of the holomorphic coupling (\ref{AQ2}) can be related and/or eliminated by the (lattice) condition that $\A(Q^2) \sim Q^2$ at $Q^2 \to 0$ and by the additional imposed condition that the difference between this coupling and its underlying (Lambert-MiniMOM) pQCD coupling $a(Q^2)$ at large $Q^2$ ($|Q^2| > {\Lambda}_{\rm L}^2$) practically disappears, namely that $\A(Q^2) - a(Q^2) \sim ({\Lambda}_{\rm L}^2/Q^2)^N$ with a large $N$ ($N=5$). This last condition represents in fact four conditions, since in general the mentioned difference is $\sim  ({\Lambda}_{\rm L}^2/Q^2)^1$. All this leads to the following five conditions:
\bes
\label{cons}
\bea
- \frac{1}{\pi} \int_{M_0^2}^{\infty} d \sigma \frac{\rho_a(\sigma)}{\sigma} &=& 
\sum_{j=1}^2 \frac{{\cal F}_j}{M_j^2} + \frac{{\cal F}_1^{(1)}}{M_1^4} + 
2 \frac{{\cal F}_1^{(2)}}{M_1^6} \ ;
\label{Q2to0}
\\
\frac{1}{\pi} \int_{-s_{L.} {\Lambda}_{\rm L}^2}^{M_0^2} d \sigma \rho_a(\sigma) &=& \sum_{j=1}^2 {\cal F}_j \ ,
\label{1u1}
\\
\frac{1}{\pi} \int_{-s_{L.} {\Lambda}_{\rm L}^2}^{M_0^2} d \sigma \sigma \rho_a(\sigma) &=& 
\sum_{j=1}^2 {\cal F}_j M_j^2 - {\cal F}_1^{(1)} \ ,
\label{1u2}
\\
\frac{1}{\pi} \int_{-s_{L.} {\Lambda}_{\rm L}^2}^{M_0^2} d \sigma \sigma^2 \rho_a(\sigma) &=& 
\sum_{j=1}^2 {\cal F}_j M_j^4 - 2 {\cal F}_1^{(1)} M_1^2 + 2 {\cal F}_1^{(2)} \ ,
 \label{1u3}
\\
\frac{1}{\pi} \int_{-s_{L.} {\Lambda}_{\rm L}^2}^{M_0^2} d \sigma \sigma^3 \rho_a(\sigma) &=& 
\sum_{j=1}^2 {\cal F}_j M_j^6 - 3 {\cal F}_1^{(1)} M_1^4 + 6 {\cal F}_1^{(2)} M_1^2 \ .
 \label{1u4}
\eea
\ees
The first of these relations represents the (lattice) condition $\A(Q^2) \sim Q^2$ when $Q^2 \to 0$.
The second relation represents the condition $\A(Q^2) - a(Q^2) <  ({\Lambda}_{\rm L}^2/Q^2)^1$, etc.,
and the last  $\A(Q^2) - a(Q^2) <  ({\Lambda}_{\rm L}^2/Q^2)^4$ (for $|Q^2| > {\Lambda}_{\rm L}^2$). In these relations, ${\Lambda}_{\rm L} =  1.153$ GeV is the (Lambert) scale appearing in Eq.~(\ref{zexpr}), and $Q_{L.}^2 \equiv s_{L.} {\Lambda}_{\rm L}^2 \approx 0.960 \ {\rm GeV}$
(with $s_{L.} = c_1^{-(c_1/\beta_0)} \approx 0.635$) represents the (Landau) branching point of the cut of the pQCD coupling $a(Q^2)$. The relations (\ref{1u1})-(\ref{1u4}) follow from the dispersion integral representation (\ref{AQ2}) of the holomorphic coupling and the analogous representation for the underlying pQCD coupling $a(Q^2)$
\be
a(Q^2) = \frac{1}{\pi}  \int_{-s_{L.} {\Lambda}_{\rm L}^2}^{\infty} d \sigma \frac{\rho_a(\sigma)}{(Q^2 + \sigma)}
\label{adisp}
\ee
where the integration over the (unphysical) cut $Q^{'2}$ ($\equiv - \sigma$) $\in [0, s_{L.} {\Lambda}_{\rm L}^2]$ is included. We refer for details to \cite{inprep}. 

In the considered coupling $\A(Q^2)$, the value of $\alpha_s(M_Z^2;\MSbar)$ will affect the underlying pQCD coupling $a(Q^2)$ and the Lambert scale ${\Lambda}_{\rm L}$, Eqs.~(\ref{aptexact})-(\ref{zexpr}), and the corresponding spectral function $\rho_a(\sigma) = {\rm Im} a(-\sigma - i \epsilon)$ appearing in Eq.~(\ref{AQ2a}). However, once $\alpha_s(M_Z^2;\MSbar)$ has been chosen, the considered coupling $\A(Q^2)$, Eqs.~(\ref{AQ2}), has altogether seven adjustable parameters: ${\cal F}_1$, ${\cal F}_1^{(1)}$, ${\cal F}_1^{(2)}$, ${\cal F}_2$, $M_1^2$, $M_2^2$ and $M_0^2$. The conditions (\ref{cons}) are five, and they can be reformulated as giving the five parameters ${\cal F}_1^{(1)}$, ${\cal F}_1^{(2)}$, ${\cal F}_2$, $M_1^2$, $M_2^2$ as functions of the two ``input'' parameters $M_0^2$ and ${\cal F}_1$. These two remaining parameters can be fixed by two additional conditions, which will be the following: (1) the coupling should achieve the local maximum at the scale $Q^2_{\ast} \approx 0.14 \ {\rm GeV}^2$, as suggested by the lattice results [after rescaling $\Lambda_{\rm MS} \mapsto \Lambda_{\rm MM}$, cf.~Fig.~(\ref{FigAlatt})]; (2) the coupling should reproduce the experimentally measured semihadronic $\tau$ decay ratio $r_{\tau} \approx 0.20$.

Since the holomorphic coupling $\A(Q^2)$ practically merges with pQCD in the high momentum regime, and the Lambert-MiniMOM renormalization scheme is not very far from $\MSbar$ scheme in the high-momentum regime,\footnote{The $\MSbar$ values (when $N_f=3$) are: $c_2 \approx 4.47$ and $c_3 \approx 20.99$. The values in the presently used 3-loop Lambert-MiniMOM scheme (and with $N_f=3$) are: $c_2=9.3$ and $c_3=48.7$.} the high momentum QCD phenomenology will be automatically reproduced. 
However, as mentioned earlier, it is important to require, in addition, that
it reproduce the measured value $r_{\tau} \approx 0.20$ \cite{ALEPH2,DDHMZ} (cf.~also App.~B of \cite{anpQCD1b}), with experimental uncertainty $\delta r_{\tau}$ less than $\pm 0.01$. Here, $r_{\tau}$ denotes the total ($V+A$) QCD massless and strangeless canonical [$r_{\tau} = a + {\cal O}(a^2)$] semihadronic decay ratio of $\tau$ lepton. This quantity is important, because it is the main low-momentum QCD quantity that is precisely measured and, simultaneously, has small higher-twist contributions (less than $\pm 0.01$ \cite{ALEPH2,DDHMZ,anpQCD1b}). Furthermore, in general the mentioned analytic QCD frameworks (\ref{Adisp}), although reproducing well the high-energy QCD phenomenology, can fail to reproduce even approximately the value of $r_{\tau} \approx 0.20$, cf.~Refs.~\cite{MSS2} in the case of APT. Although this quantity is a timelike quantity, it can be expressed via the use of the Cauchy theorem in terms of the Adler function $d(Q^2)$ which is a spacelike quantity
\cite{Braaten1,Braaten2,Pivovarov:1991rh,LeDiberder:1992te}:
\bes
\label{rtaucont}
\bea
r_{\tau} &=& \frac{1}{2 \pi} \int_{-\pi}^{+ \pi}
d \phi \ (1 + e^{i \phi})^3 (1 - e^{i \phi}) \
d(Q^2=m_{\tau}^2 e^{i \phi}) \ ,
\label{rtauconta}
\\
r_{\tau}(D=0) &=& \frac{1}{2 \pi} \int_{-\pi}^{+ \pi}
d \phi \ (1 + e^{i \phi})^3 (1 - e^{i \phi}) \
d(Q^2=m_{\tau}^2 e^{i \phi} D=0) \ ,
\label{rtauD0cont}
\eea
\ees
where the canonical massless Adler function $d(Q^2)$ is a derivative of the
quark current correlator, $d (Q^2) = - d \Pi(Q^2)/d \ln Q^2$, whose leading-twist (dimension $D=0$) perturbation theory expansion
\be
d(Q^2; D=0)_{\rm pt} =  a(Q^2) + \sum_{n=1}^{3} d_n a(Q^2)^{n+1}  + {\cal O}(a^5)
\label{dpt}
\ee
 is known to $\sim a^4$ \cite{d1,d2,d3}. The small higher-dimension (higher-twist) contributions in $r_{\tau}$ come from higher-dimension contributions of the Adler function. In Eq.~(\ref{rtauD0cont}) we denoted the leading-twist (dimension $D=0$) parts of these quantities. We note that it is expected that $r_{\tau}(D=0) \approx r_{\tau} \approx 0.20$ [cf.~the next Section, Eqs.~(\ref{rtauexp})]. With the holomorphic coupling $\A(Q^2)$, which does not have the perturbative running, the truncated series (\ref{dpt}) can be evaluated by replacing the powers $a(Q^2)^{n+1}$ by their holomorphic analogs $\A_{n+1}(Q^2)$ [$\not= \A(Q^2)^{n+1}$] constructed entirely from $\A(Q^2)$ [$\equiv \A_1(Q^2)$], as a linear combination of the logarithmic derivatives of $\A(Q^2)$, according to the procedure of Ref.~\cite{CV12}\footnote{$\A_{n+1}(Q^2)$ is a linear combination of $\tA_{k+1}(Q^2) \propto (d/d \ln Q^2)^k \A(Q^2)$ with $k=n,n+1,\ldots, N_{\rm max}$, where in our considered case $N_{\rm max}=3$. If $n$ is noninteger, the construction of $\tA_{k+1}(Q^2)$ and $\A_{n+1}(Q^2)$ from $\A(Q^2)$ was presented in \cite{GCAK}.}
\bes
\label{dan}
\bea
d(Q^2;D=0)_{\rm an} &=&  \A(Q^2) + \sum_{n=1}^{3} {\widetilde d}_n \tA_{n+1}(Q^2)  + {\cal O}(\tA_5) 
\label{dana}
\\
&=&  \A(Q^2) + \sum_{n=1}^{3} d_n \A_{n+1}(Q^2)  + {\cal O}(\A_5) \ .
\label{danb}
\eea
\ees
Here, for simplicity, the renormalization scale $\mu^2 \equiv \kappa Q^2$ was set equal to $Q^2$ (i.e., $\kappa=1$).\footnote{
Unlike $d_1={\bar d}_1$, the coefficients $d_2$ and $d_3$ are scheme ($c_2, c_3$) dependent: $d_2 = {\bar d}_2 - (c_2 - {\bar c}_2)$, and $d_3 = {\bar d}_3 - 2 {\bar d}_1 (c_2 - {\bar c}_2) - (1/2)(c_3 - {\bar c}_3)$, where bars denote the quantities in the $\MSbar$ scheme.}
This truncated series can be efficiently resummed in any holomorphic coupling framework, by using a generalization  \cite{BGApQCD1,BGApQCD2,BGA,anOPE} of the diagonal Pad\'e resummation approach \cite{GardiPA}, generalized in such a way that it gives an exactly renormalization scale independent result
\be
d(Q^2;D=0)_{\rm res} = 
\tal_1 \; \A_1(\kappa_1 Q^2)+ (1-\tal_1) \; \A_1(\kappa_2 Q^2)  +  {\cal O}(\A_5) \ .
\label{dBGan22}
 \ee
 with $\tal_1$, $\kappa_1$, $\kappa_2$ being positive parameters obtained uniquely from the known perturbation coefficients $d_1$, $d_2$ and $d_3$ (using any renormalization scale):\footnote{
   In Ref.~\cite{BGApQCD1} it was demonstrated that the result is exactly independent of the renormalization scheme. In \cite{BGApQCD2} the method was extended to truncated perturbation series with uneven number of terms. In Refs.~\cite{BGA,anOPE} this method was revived and applied to QCD frameworks with holomorphic (analytic) couplings, where it was shown that it works remarkably well, due to the absence of Landau singularities in the coupling. In Refs.~\cite{BGA,Techn} it was shown that in holomorphic QCD frameworks the method gives a convergent sequence when the number of terms in the perturbation series is increased, thus eliminating the renormalon ambiguity in such frameworks.}
$\tal_1 \approx 1.14642$, $\kappa_1 \approx 0.66453$, $\kappa_2 \approx 5.91622$.
It turns out that the application of the two methods (\ref{dan})-(\ref{dBGan22}) leads in Eq.~(\ref{rtaucont}) to almost the same values of $r_{\tau}$, which indicated a good stability of evaluation of  $r_{\tau}$ in the present framework of $\A(Q^2)$ coupling. In \cite{BenekeJamin,Caprini} the Adler function $d(Q^2; D=0)$ was evaluated, i.e., resummed, by other methods, using pQCD coupling. 

To recapitulate, the constructed coupling has seven free parameters:\footnote{We recall that (Lambert) ${\Lambda}_{\rm L}$ scale was fixed by the world average value of the underlying pQCD coupling $a$, or equivalently, by the value of $\alpha_s(M_Z^2;\MSbar)$.} ${\cal F}_1$, ${\cal F}_2$, ${\cal F}_1^{(1)}$, ${\cal F}_1^{(2)}$, $M_1^2$, $M_2^2$, $M_0^2$. The five conditions (\ref{cons}) allow  us to reduce the number of free parameters to two, which we chose to be the coefficient ${\cal F}_1$ and the (pQCD-onset)  scale $M_0^2$. These two free parameters were then adjusted so that the correct value of $r_{\tau}$ was reproduced, and that $\A(Q^2)$ acquired the maximum at $Q^2_{\ast} \approx 0.14 \ {\rm GeV}^2$ as suggested by lattice calculations, cf.~Fig.~\ref{FigAlatt} (we recall that the lattice results have been rescaled in Fig.~\ref{FigAlatt}  back to the usual $\MSbar$ scaling, i.e., $\Lambda_{\rm MM} \mapsto \Lambda_{\MSbar}$). This then allowed us to determine the parameters ${\cal F}_1 \approx 0.0454 {\Lambda}_{\rm L}^2$ and $M_0^2 \approx 3 {\Lambda}_{\rm L}^2$ [in the considered central case, i.e., when $\alpha_s(M_Z^2;\MSbar)=0.1185$]. The resulting parameters, which fully determine the sought for holomorphic nearly perturbative coupling $\A(Q^2)$, are given in Table \ref{tabres}, first line. They are expressed in dimensionless form: $f_j \equiv {\cal F}_j/{\Lambda}_{\rm L}^2$ ($j=1,2$), $f_1^{(1)} \equiv  {\cal F}_1^{(1)}/{\Lambda}_{\rm L}^4$, $f_1^{(2)} \equiv  {\cal F}_1^{(2)}/{\Lambda}_{\rm L}^6$; $s_j = M_j^2/{\Lambda}_{\rm L}^2$ ($j=1,2,3$). 
 \begin{table}
   \caption{The (dimensionless) parameters of the coupling $\A(Q^2)$: $s_j \equiv M_j^2/{\Lambda}_{\rm L}^2$ ($j=0,1,2$); $f_j \equiv {\cal F}_j/{\Lambda}_{\rm L}^2$ ($j=1,2$); $f_1^{(1)} \equiv {\cal F}_1^{(1)}/{\Lambda}_{\rm L}^4$; $f_1^{(2)} \equiv {\cal F}_1^{(2)}/{\Lambda}_{\rm L}^6$. Included is the Lambert scale ${\Lambda}_{\rm L}$ (in GeV). The parameters are adjusted so that $r_{\tau}(D=0) \approx 0.201$ when the resummed expression (\ref{dBGan22}) is used in $r_{\tau}(D=0)$, and that the maximal value of $\A(Q^2)$ is achieved at $Q_{\ast}^2 \approx 0.14 \ {\rm GeV}^2$. Three cases of the input parameter $\alpha_s(M_Z^2;\MSbar)$ are given. Our central case will be the first line, $\alpha_s(M_Z^2;\MSbar)=0.1185$.}
\label{tabres}  
\begin{ruledtabular}
\begin{tabular}{l|ll|llllll}
$\alpha_s(M_Z^2;\MSbar)$ &  $s_0$ & $f_1$  & $f_1^{(1)}$ & $f_1^{(2)}$ & $s_1$ & $f_2$ & $s_2$ & ${\Lambda}_{\rm L}$ [GeV]
  \\
  \hline
$0.1185$ & $3.000$ & $0.04537$ & $+1.880 \times 10^{-3}$ &  $- 2.399 \times 10^{-4}$ & $0.07952$ & $0.02687$ & $2.1518$ &  $1.5130$
\\
\hline
$0.1181$ & $3.522$ & $0.04937$ & $-2.681 \times 10^{-3}$ &  $- 1.294 \times 10^{-5}$ & $0.05774$ & $0.02913$ & $2.5549$ &  $1.4862$
\\
$0.1189$ & $2.542$ & $0.04108$ & $+5.223 \times 10^{-3}$ &  $-3.919 \times 10^{-4}$ & $0.08411$ & $0.02539$ & $1.7859$ &  $1.5400$
\end{tabular}
\end{ruledtabular}
\end{table}
 The value of $Q^2=Q_{\ast}^2$ where the maximum of $\A(Q^2)$ is reached is in this case $Q_{\ast}^2 = 0.141 \ {\rm GeV}^2$, and the maximal value is $\A(Q_{\ast}^2) = 1.040/\pi$. The value of $r_{\tau}(D=0)$ is $0.203$ if determined in Eq.~(\ref{rtauD0cont}) by the method (\ref{dan}), and $0.201$ if by the method (\ref{dBGan22}), both consistent with the measured value $r_{\tau} \approx 0.20$ [$\approx r_{\tau}(D=0)$] -- for the measured values, cf.~the next Section, Eqs.~(\ref{rtauexp}).

The truncated series (\ref{dan}), when inserted into the contour integral (\ref{rtaucont}), gives the following series: $r_{\tau}(D=0) = 0.153 + 0.049 - 0.009 + 0.011 = 0.153 +0.026 + 0.003 + 0.021$ ($\approx 0.203$), where the first series corresponds to the sum (\ref{dana}) and the second to the ``reorganized'' sum (\ref{danb}). The $\MSbar$ pQCD approach gives, on the other hand, the analogous results $r_{\tau}(D=0) = 0.138+0.032+0.008+0.004 = 0.138+0.026+0.010+0.007$ ($\approx 0.182$). Incidentally, we can see that $\MSbar$ approach gives for this ($D=0$) $\tau$-decay ratio a result which is surprisingly far from the expected value $r_{\tau}(D=0) \approx 0.20$. We will comment on this problem later on [after Eqs.~(\ref{rtauexp})]. If using pQCD in 3-loop Lambert-MiniMOM scheme [Eq.~(\ref{aptexact}) with $c_2=9.3$, i.e., the pQCD coupling $a$ which is underlying the holomorphic coupling $\A$], we obtain similarly low value $r_{\tau}(D=0) \approx 0.184$.

\begin{figure}[htb] 
\centering\includegraphics[width=100mm]{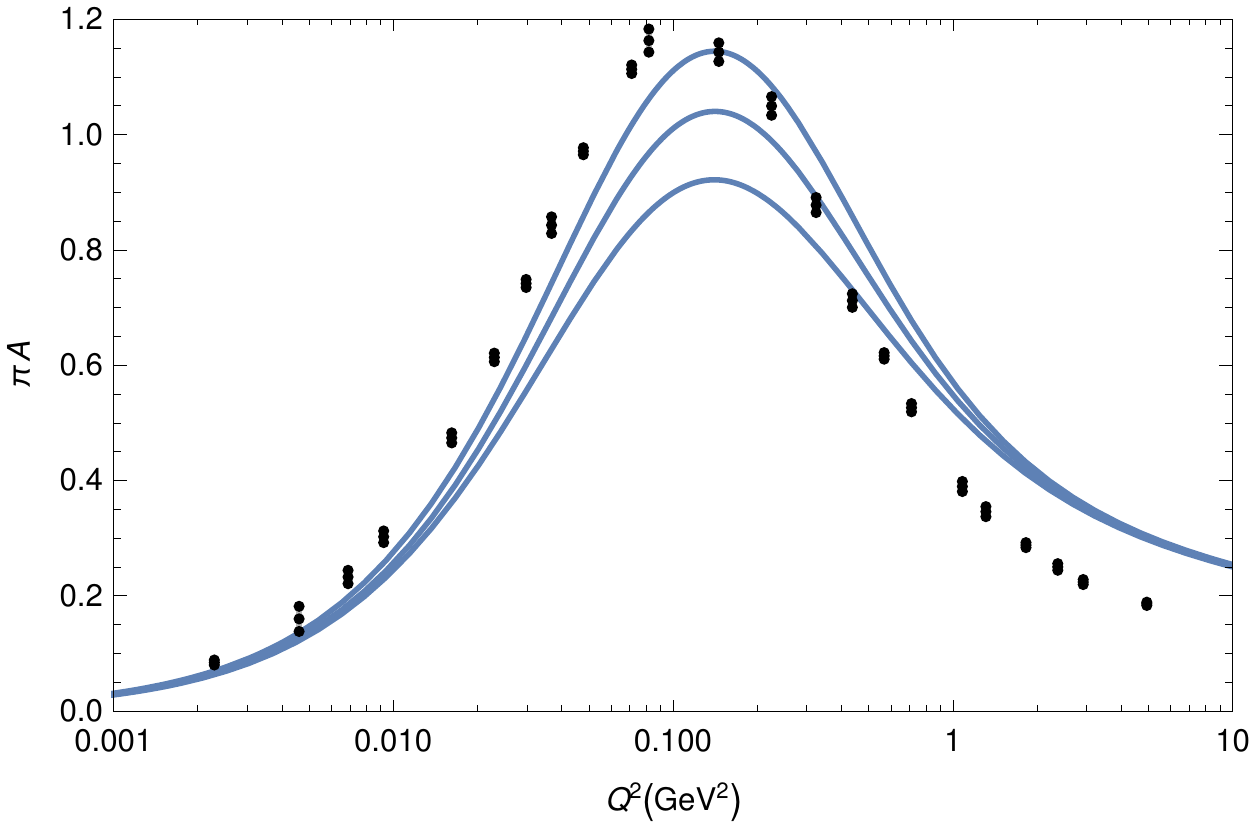}
 \caption{\footnotesize  The constructed $\A$QCD coupling $\pi \A(Q^2)$, at low positive $Q^2$, for $N_f=3$ and in the Lambert-MiniMOM scheme, for the three different cases of Table \ref{tabres}: $\alpha_s(M_Z^2;\MSbar)=0.1185$ and $\alpha_s(M_Z^2;\MSbar)=0.1185 \pm 0.0004$. The nonperturbative (lattice) values $\pi \A_{\rm latt.}(Q^2)$ of Ref.~\cite{LattcoupNf0} are included as points, as in Fig.~\ref{FigAlatt}.}
\label{FigLatt3A}
\end{figure}
In Table \ref{tabres} we also included, for comparison, the results for the coupling when the values $\alpha_s(M_Z^2;\MSbar)=0.1185 \pm 0.0004$ are taken (the second and third lines). The resulting coupling $\A(Q^2)$ in the three cases, and the lattice results, at positive $Q^2$, are given in Fig.~\ref{FigLatt3A}.\footnote{Particle Data Group in 2014 gives the world average ${\overline a}(M_Z^2)=(0.1185 \pm 0.0006)/\pi$ \cite{PDG2014}; in 2016 it gives ${\overline a}(M_Z^2)=(0.1181 \pm 0.0011)/\pi$ \cite{PDG2016}.}

 The values of this coupling, for low positive values of $Q^2$, for the central case $\alpha_s(M_Z^2;\MSbar)=0.1185$ were already presented in Fig.~\ref{FigAlatt} as the continuous line, together with the (rescaled) values from the lattice calculation. As explained earlier, at low $Q^2 \alt 1 \ {\rm GeV}^2$, we should not expect a quantitative agreement between $\A(Q^2)$ and the lattice coupling, but only a qualitative agreement. At higher $Q^2 \agt 10^1 \ {\rm GeV}^2$, the agreement should in principle be better. Nonetheless, the lattice calculations in Fig.~\ref{FigAlatt} were performed in the quenched ($N_f=0$) approximation, while our coupling $\A(Q^2)$ has $N_f=3$ (which is the realistic choice for the regime $10^{-2} \ {\rm GeV}^2 \lesssim Q^2 \lesssim 10^1 \ {\rm GeV}^2$). We checked that this effect is responsible for about one third of the difference between $\A(Q^2)$ and the lattice coupling for $Q^2 \sim 10^0$-$10^1 \ {\rm GeV}^2$  in Fig.~\ref{FigAlatt}. However, the main reason for the difference in the regime $Q^2 \sim 10^0$-$10^1 \ {\rm GeV}^2$ appears to reside in the following: the lattice results in Fig.~\ref{FigAlatt} (from Ref.~\cite{LattcoupNf0}) are good, i.e., are close to the continuum limit, only for deep IR regime $Q^2 < 1 \ {\rm GeV}^2$, but for higher $Q^2$ they suffer from so called hypercubic lattice artifacts \cite{SternbeckComm} due to the lattice being coarse ($\beta = 5.7$).

\begin{figure}[htb] 
\centering\includegraphics[width=100mm]{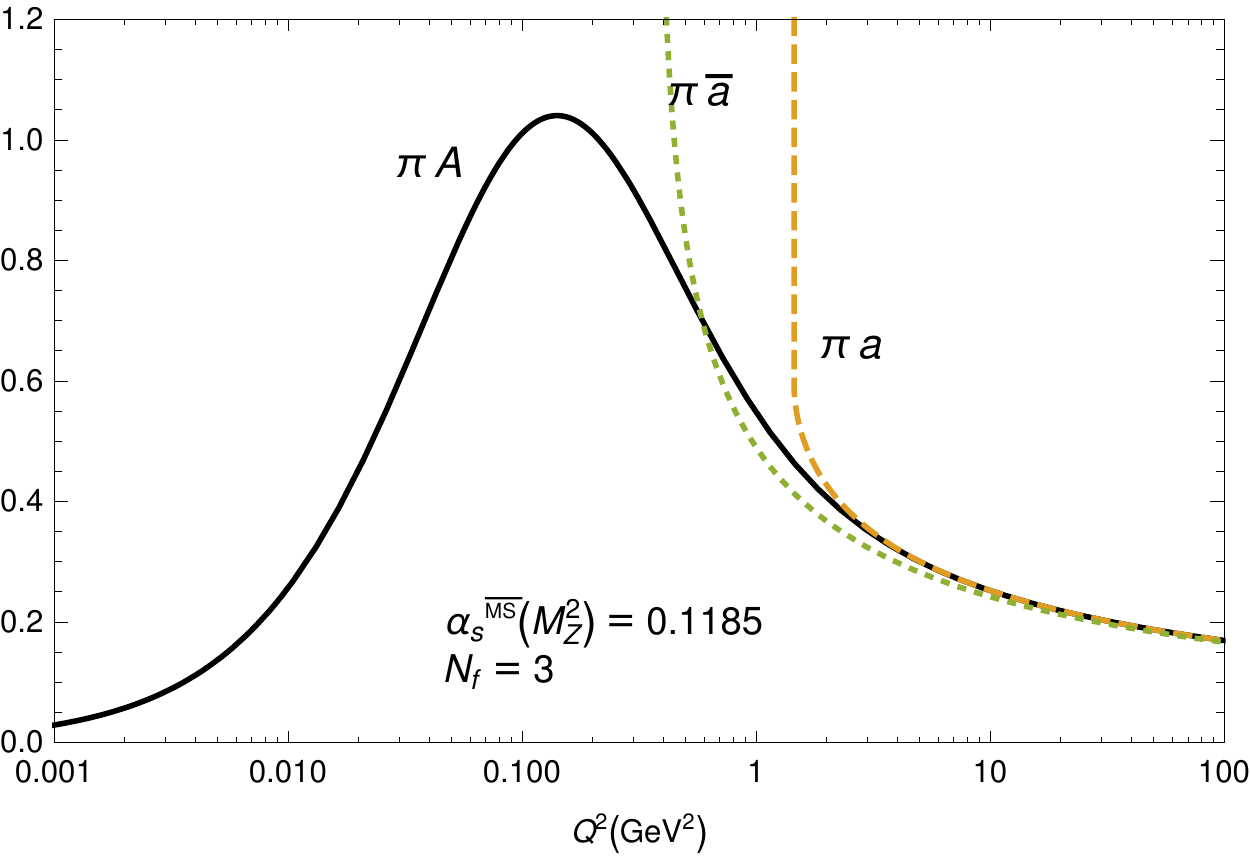}
\caption{\footnotesize  The constructed (nearly perturbative) holomorphic coupling $\pi \A(Q^2)$ (continuous line) as compared to its perturbative analog $\pi a(Q^2)$ (dashed line), in the Lambert-MiniMOM scheme. The two couplings practically merge at $Q^2 > 4 \ {\rm GeV}^2$. For further comparison, the usual $\MSbar$ coupling $\pi {\overline a}(Q^2)$ is included.}
\label{FigAa}
 \end{figure}

In Fig.~\ref{FigAa} we show, for positive $Q^2$ values, the comparison between the constructed coupling $\pi \A(Q^2)$ and the corresponding pQCD coupling $\alpha_s(Q^2) \equiv \pi a(Q^2)$ Eq.~(\ref{aptexact}), as well as the usual $\MSbar$ pQCD coupling $\alpha(Q^2,\MSbar) \equiv \pi {\overline a}(Q^2)$, all for $N_f=3$.
 The two couplings $\pi \A(Q^2)$ and $\pi a(Q^2)$ practically merge with each other at higher $Q^2$ values, in accordance with the conditions (\ref{1u1})-(\ref{1u4}). For example, at $Q^2 = 4 \ {\rm GeV}^2$ the relative difference between them, ($a/\A - 1$), is about $5 \times 10^{-3}$, and at $Q^2 = 6 \ {\rm GeV}^2$ it is about $1 \times 10^{-3}$. The Landau branching point of the coupling $a(Q^2)$ in Lambert-MiniMOM scheme is at $Q^2 \approx 1.453 \ {\rm GeV}^2$ ($Q \approx 1.205$ GeV; it is not a pole), while for the $\MSbar$ coupling  ${\overline a}(Q^2)$ it is at $Q^2 \approx 0.371 \ {\rm GeV}^2$ ($Q \approx 0.609$ GeV; it is a pole).

We stress that both $\A(Q^2)$ and $a(Q^2)$ represent the running coupling in the same 3-loop Lambert-MiniMOM renormalization scheme in the perturbative sense (i.e., $c_2=9.3$; $c_3=48.7$; etc.), and they are practically the same for $|Q^2| > {\Lambda}_{\rm L}^2$. Nonetheless, $\A(Q^2)$ fulfills, in contrast to $a(Q^2)$, a host of attractive properties: (a) at $0 < Q^2 < 1 \ {\rm GeV}^2$ it behaves qualitatively as suggested by large-volume lattice calculations; (b) it reproduces well the measured value of semihadronic $\tau$ decay ratio $r_{\tau}$; (c) it has no Landau ghosts, namely, it is holomorphic for all $Q^2 \in \mathbb{C} \backslash (-\infty, -M_{\rm thr}^2]$ where the square of the threshold mass is by construction positive ($M_{\rm thr}^2 = M_1^2 = s_1 {\Lambda}_{\rm L}^2 \approx 0.182 \ {\rm GeV}^2$).
 
 \section{Application of Borel sum rules for $V+A$ spectral functions}
 \label{sec:SR}

We point out that we consider the discussed coupling $\A(Q^2)$ as universal, and we consider as universal the expectation values $\langle O_D \rangle$ of operators appearing in the Operator Product Expansion (OPE) for inclusive observables. Technically speaking, since the difference between $\A(Q^2)$ and its underlying pQCD coupling $a(Q^2)$ is by construction $\sim ({\Lambda}_{\rm L}^2/Q^2)^5$ (for $|Q^2| > {\Lambda}_{\rm L}^2$), the dimensionality $D$ of operators that can be applied in OPE unambiguously has upper bound $D < 10$, i.e., $D_{\rm max} = 8$.

\begin{figure}[htb] 
\centering\includegraphics[width=100mm]{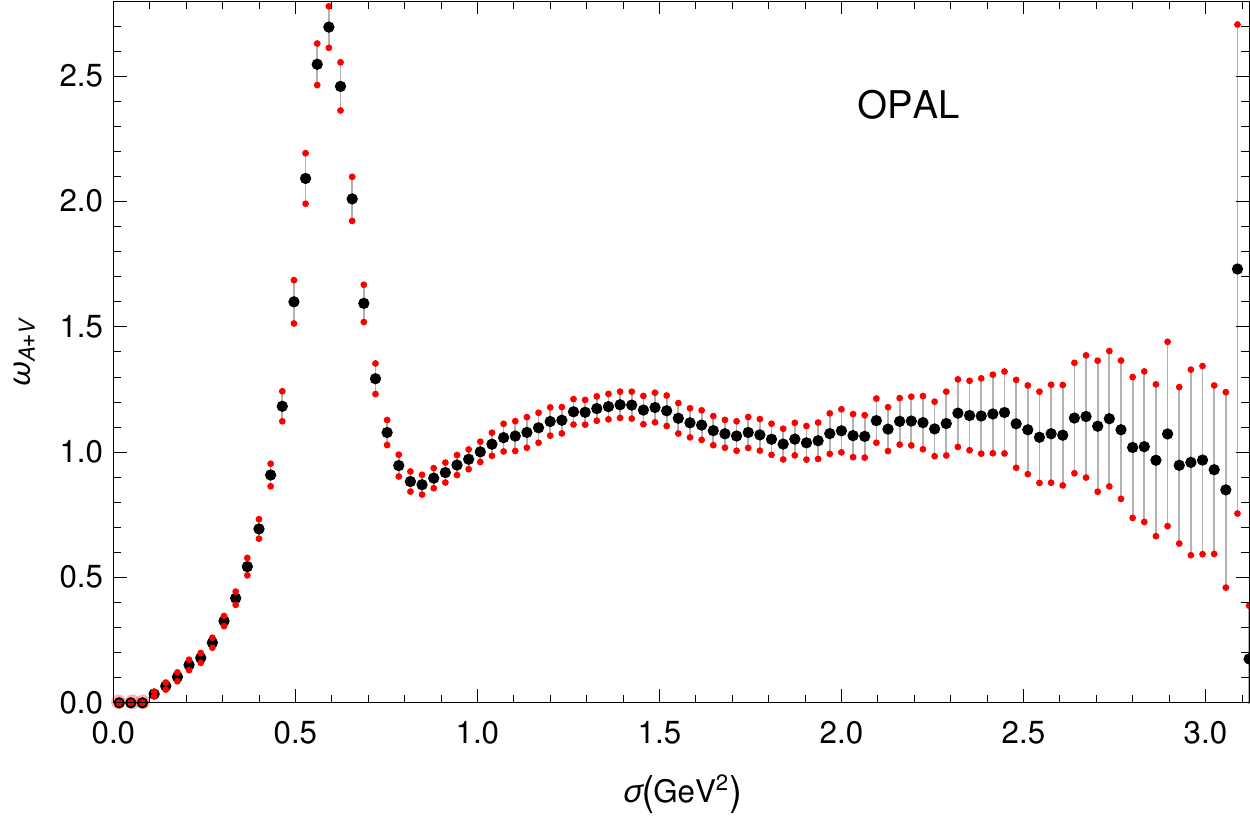}
\caption{\footnotesize  The total spectral function $\omega \equiv \omega_{V+A}$ as measured by OPAL \cite{OPAL,PerisPC1}. The pion peak contribution $2 \pi^2 f_{\pi}^2 \delta(\sigma-M_{\pi}^2)$ (where $f_{\pi}=0.1305$ GeV) must be included but is not visible in the Figure because extremely narrow.}
\label{FigOPAL}
 \end{figure}

Here we briefly present an application of the obtained coupling to the Borel sum rules \cite{Ioffe1,Ioffe2} for the $\tau$ decay $V+A$ spectral functions as measured by OPAL \cite{OPAL,PerisPC1}.\footnote{
  We are grateful to S.~Peris for providing us with the measured spectral functions and covariance matrices of OPAL Collaboration; these data are the update, made by the authors of Ref.~\cite{PerisPC1}, of the OPAL data, based on the older OPAL data given to them by S.~Menke.}
A more detailed analysis will be presented elsewhere \cite{inprep}. These sum rules are based on the application of the Cauchy theorem to the ($V+A$) quark current correlator function $\Pi(Q^2)$
 \be
\int_0^{\sigma_{0}} d \sigma g(-\sigma) \omega_{\rm exp}(\sigma)  =
-i \pi  \oint_{|Q^2|=\sigma_0} d Q^2 g(Q^2) \Pi_{\rm th}(Q^2)  \ ,
\label{sr1}
\ee
where the spectral function  $\omega(\sigma)$ is proportional to the discontinuity of the (otherwise holomorhic) current correlator across its cut
\be
\omega(\sigma) \equiv 2 \pi \; {\rm Im} \ \Pi_{V+A}(Q^2=-\sigma - i \epsilon) \ ,
\label{om1}
\ee
where $\Pi_{V+A}= \Pi^{(0+1)}_{ud,V}+\Pi^{(0+1)}_{ud,A}$ (notations of Refs.~\cite{Braaten2}), and $g(Q^2)$ is any function holomorphic in the entire complex $Q^2$-plane, and $\sigma_{0}$ ($\leq m_{\tau}^2$) is a chosen upper bound. The choice of function $g(Q^2)$ specifies the sum rule. The spectral functions $\omega(\sigma)_A$ and $\omega(\sigma)_V$ have been measured \cite{OPAL}, and the resulting values of the total spectral function $\omega(\sigma) \equiv \omega(\sigma)_{A+V}$ are given in Fig.~\ref{FigOPAL}.
 The right-hand side of the sum rule (\ref{sr1}) is evaluated theoretically, with the correlator function determined by the OPE approach
 \be
\Pi_{V+A}(Q^2) =  - \frac{1}{2 \pi^2} \ln(Q^2/\mu^2) + 
\Pi(Q^2;D\!=\!0)
+ \sum_{n \geq 2} \frac{ \langle O_{2n} \rangle_{V+A}}{(Q^2)^n} \left(
1 + {\cal C}_n a(Q^2) \right) \ .
\label{OPE1}
\ee
The terms ${\cal C}_n a(Q^2)$ in $D$-dimensional contribution ($D=2 n \geq 4$) turn out to be negligible as well as the $D=2$ terms ($\sim {\cal O}(m^2_{u.d})$) \cite{Ioffe1,Ioffe2,PerisPC1}. 

In the following we specifically concentrate on the Borel sum rules, which are defined by choosing the function $g(Q^2)$ to be
\be
 g(Q^2) \equiv g_{M^2}(Q^2) = \frac{1}{M^2} \exp( Q^2/M^2) \ ,
 \label{gQ2}
 \ee
 where $M^2$ is an arbitrarily chosen complex scale with ${\rm Re}(M^2)>0$. It is convenient to perform now integration by parts on the right-hand side of Eq.~(\ref{sr1}), leading in the Borel case (\ref{gQ2}) to the following sum rules:
\bea
\frac{1}{M^2} \int_0^{\sigma_0} d \sigma \exp( - \sigma/M^2) \omega_{\rm exp}(\sigma) &=&
- \frac{i}{2 \pi}  \int_{\phi=-\pi}^{\pi} \frac{d Q^2}{Q^2}
{\cal D}_{V+A}(Q^2) \left[ \exp( Q^2/M^2) -  \exp( -\sigma_0/M^2) \right] {\big |}_{Q^2 = \sigma_0 \exp(i \phi)} .
\label{sr2}
\eea
Here, ${\cal D}(Q^2)$ is the (full) massless Adler function
\bea
{\cal D}_{V+A}(Q^2) &\equiv&  - 2 \pi^2 \frac{d \Pi_{V+A}(Q^2)}{d \ln Q^2} 
 =  1 + d(Q^2;D=0) + 2 \pi^2 \sum_{n \geq 2}
 \frac{ n \langle O_{2n} \rangle_{V+A}}{(Q^2)^n}  \ ,
\label{Adlfull}
\eea
where the OPE expression (\ref{OPE1}) was used, without the negligible terms $\propto {\cal C}_n$. The central objective in the theoretical evaluation is the evaluation of the $D=0$ (leading-twist) Adler function $d(Q^2) = - 2 \pi^2 d \Pi(Q^2,D=0)/ d \ln Q^2$, whose perturbation expansion is given in Eq.~(\ref{dpt}), and can be evaluated with our (nearly perturbative) holomorphic coupling either in the truncated form (\ref{dan}) or in the resummed form (\ref{dBGan22}). More explicitly, the Borel sum rule for the real part has the form
\be
{\rm Re} B_{\rm exp}(M^2) =  {\rm Re} B_{\rm th}(M^2) \ ,
\label{sr3a}
\ee
where
\bes
\label{sr3}
\bea
B_{\rm exp}(M^2) &\equiv& \int_0^{\sigma_{0}} 
\frac{d \sigma}{M^2} \; \exp( - \sigma/M^2) \omega_{\rm exp}(\sigma)_{V+A} \ ,
\label{sr3b}
\\
B_{\rm th}(M^2) &\equiv&  \left( 1 - \exp(-\sigma_{0}/M^2) \right)
+ B(M^2;D\!=\!0) + 2 \pi^2 \sum_{n \geq 2}
 \frac{ \langle O_{2n} \rangle_{V+A}}{ (n-1)! \; (M^2)^n} \ ,
\label{sr3c}
\eea
\ees
The $D=0$ part is\footnote{Contour integrals of $D=0$ part  $d(Q^2)$ of the Adler function, but with polynomial weight functions $W_i(Q^2/\sigma_0)$, were studied within pQCD in Refs.~\cite{BenekeJamin}.}
\bea
B(M^2;D\!=\!0) &=&
\frac{1}{2 \pi}\int_{-\pi}^{\pi}
d \phi \; d(Q^2\!=\!\sigma_{0} e^{i \phi};D=0) \left[ 
\exp \left( \frac{\sigma_{0} e^{i \phi}}{M^2} \right) -
\exp \left( - \frac{\sigma_{0}}{M^2} \right) \right] \ .
\label{BD0}
\eea
We keep only the $D=4$ and $D=6$ terms in the OPE ($D=2$ term is negligible). The attractive advantages of the Borel sum rules are:
\begin{enumerate}
\item
  For the low scales $M^2$ the Borel transform $B(M^2)$ is probing the low-momentum regime (low $\sigma$). In general, the high-momentum regime contributions, where experimental uncertainties are higher, are suppressed in $B(M^2)$.
  \item
  If $M^2 = |M^2| \exp(i \pi/6)$, it can be checked that the $D=6$ term in ${\rm Re} B_{\rm th}(M^2)$ is zero. If $M^2 = |M^2| \exp(i \pi/4)$ it turns out that the $D=4$ term in ${\rm Re} B_{\rm th}(M^2)$ is zero.
\end{enumerate}

Therefore, we may determine the $\langle O_{4} \rangle_{V+A}$ condensate by comparing ${\rm Re} B_{\rm th}(M^2)$  with ${\rm Re} B_{\rm exp}(M^2)$ at  $M^2 = |M^2| \exp(i \pi/6)$; and the $\langle O_{6} \rangle_{V+A}$ condensate when comparing at $M^2 = |M^2| \exp(i \pi/4)$. Subsequent application of the Borel sum rule $B(M^2)$ for real $M^2 >0$, where both $\langle O_{4} \rangle_{V+A}$ and $\langle O_{6} \rangle_{V+A}$ contribute to $B_{\rm th}(M^2)$, then gives a prediction whose quality can be judged by comparing with $B_{\rm exp}(M^2)$.

The gluon condensate $\langle a GG \rangle$ is related to $\langle O_{4} \rangle_{V+A}$. Namely, in the (justified) approximation of neglecting the terms ${\cal O}(m_{u,d}^4)$ \cite{PerisPC1}, the two condensates $\langle O_{4} \rangle_{V}$  and $\langle O_{4} \rangle_{A}$ are equal to each other, and we have \cite{Braaten2}
\bes
\label{aGGO4}
\bea
\langle O_{4} \rangle_{V+A} &=& \frac{1}{6} \langle a GG \rangle + 2 (m_u + m_d) \langle {\bar q} q \rangle \; \Rightarrow
\label{aGGO4a}
\\
\langle a GG \rangle &=& 6 \langle O_{4} \rangle_{V+A} - 12 (m_u + m_d) \langle {\bar q} q \rangle
= 6 \langle O_{4} \rangle_{V+A} + 6 f_{\pi}^2 m_{\pi}^2 \approx 6 \langle O_{4} \rangle_{V+A} + 0.0020 \ {\rm GeV}^4.
\label{aGGO4b}
\eea
\ees
Here, we neglected corrections of relative order ${\cal O}(a)$, we denoted as $\langle {\bar q} q \rangle$ the condensate $\langle {\bar u} u \rangle \approx \langle {\bar d} d \rangle$ \cite{Ioffe1,Ioffe2,PerisPC1}, and used the PCAC relation \cite{PCAC} with the values $f_{\pi} = 0.1305$ GeV and $m_{\pi} = 0.13957$ GeV \cite{PDG2016}.

\begin{figure}[htb] 
\begin{minipage}[b]{.49\linewidth}
  \centering\includegraphics[width=85mm]{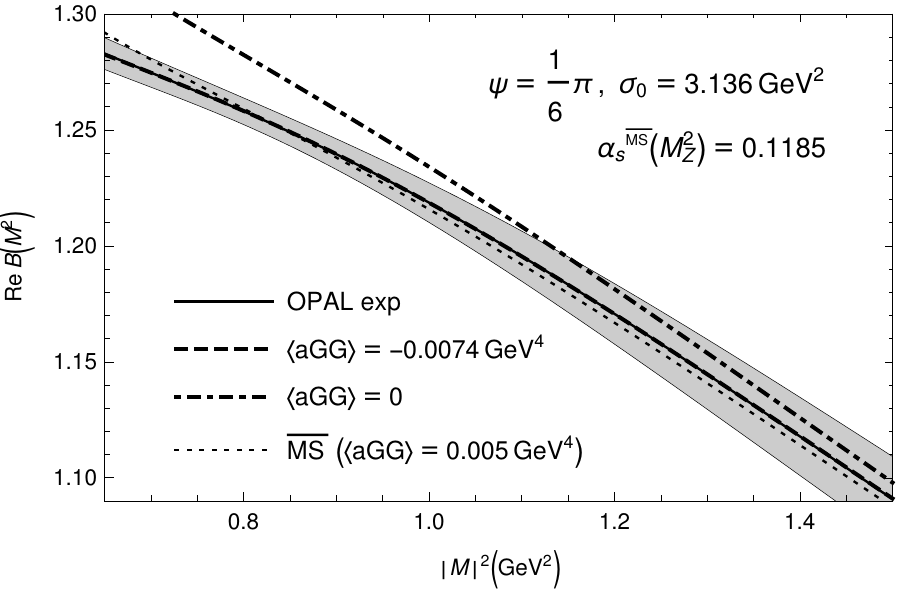}
  \end{minipage}
\begin{minipage}[b]{.49\linewidth}
  \centering\includegraphics[width=85mm]{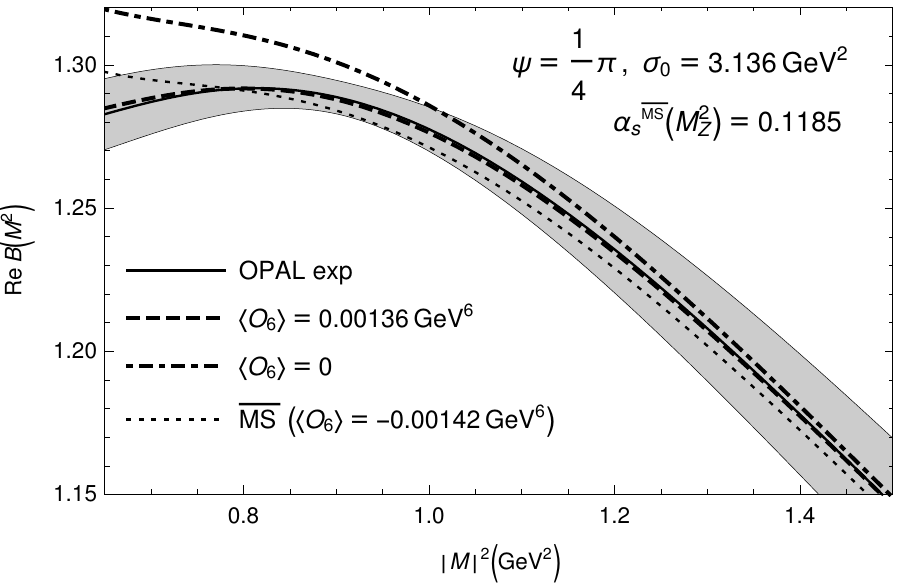}
\end{minipage}
\caption{\footnotesize Borel transforms ${\rm Re} B(M^2)$ for $\Psi \equiv {\rm Arg}(M^2) = \pi/6$ (left-hand side) and $\Psi = \pi/4$ (right-hand side), as a function of $|M^2|$. The experimental results are presented as grey band. The central experimental curve (grey line) is in both Figures almost covered by the theoretical curve (dashed line) of the holomorphic QCD (AQCD), with the fitted value $\langle a G G \rangle = - 0.0099 \ {\rm GeV}^4$ (left-hand side) and $\langle O_{6} \rangle_{V+A} = +0.0014 \ {\rm GeV}^6$ (right-hand side). For comparison, the AQCD curve with zero values of condensates is included (dot-dashed), as well as the fitted theoretical curve of $\MSbar$ pQCD (dotted). The minimized $\chi^2$ values are in $\Psi=\pi/6$ case: $1.3 \times 10^{-8}$ for AQCD and $1.4 \times 10^{-5}$ for $\MSbar$; and  in $\Psi=\pi/4$ case: $9.6 \times 10^{-7}$ for AQCD and $3.8 \times 10^{-5}$ for $\MSbar$.}
\label{FigPi64}
\end{figure}

\begin{figure}[htb] 
\centering\includegraphics[width=120mm]{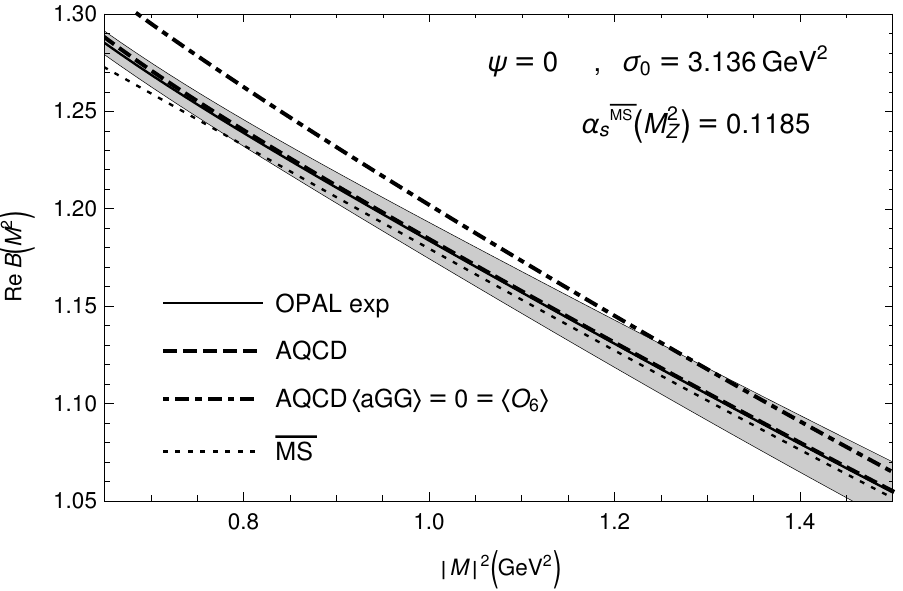}
\caption{\footnotesize  The resulting $B(M^2)$ for real positive $M^2$, with the condensate values as determined by the fit presented in Figs.~\ref{FigPi64}. The experimental results are presented as grey band. The central experimental curve (grey) is almost covered by the theoretical curve of the holomorphic QCD (AQCD) case (dashed), with the condensate values determined in Figs.~\ref{FigPi64}. Included is also AQCD curve with zero condensates (dot-dashed), and $\MSbar$ curve with the condensate values determined in Figs.~\ref{FigPi64}.
The resulting values of $\chi^2$ values are $1.3 \times 10^{-6}$ in AQCD and $2.8 \times 10^{-5}$ in $\MSbar$.}
\label{FigPsi0}
 \end{figure}

In Figs.~\ref{FigPi64} we present the results for the parameters  $\langle a G G \rangle$ and  $\langle O_{6} \rangle_{V+A}$ condensates, obtained by fitting the theoretical curves to the central experimental curves, for our framework with $\A(Q^2)$ and for the usual pQCD framework in $\MSbar$. The fitting was performed, for simplicity, by the least squares method with constant weights.
For the upper integration bound in the sum rules we used the maximal value of $\sigma$ in the OPAL bins $\sigma_0 = 3.136 \ {\rm GeV}^2$ (this is somewhat lower than $m_{\tau}^2=3.1572 \ {\rm GeV}^2$). In our framework we evaluated the Adler function with the resummation method (\ref{dBGan22})\footnote{If using the truncated evaluation method (\ref{dan}), the results differ only insignificantly.}, and in $\MSbar$ case we evaluated the truncated perturbation series (\ref{dpt}). Finally, in Fig.~\ref{FigPsi0} we present the resulting curves when ${\rm Arg}(M^2) = 0$;
we can see that the resulting curve in the case of the holomorphic coupling $\A(Q^2)$  (AQCD) is significantly better than in the $\MSbar$ pQCD case. The fitted values, obtained by the method of minimal $\chi^2$, are\footnote{
 The (minimized) $\chi^2$ values were evaluated as an averaged equal weight sum of squared deviations, $\chi^2(\Psi) = (1/n) \sum_{\alpha=0}^n ( {\rm Re} B_{\rm th}(M_{\alpha}^2) - {\rm Re} B_{\rm exp}(M_{\alpha}^2) )^2$, where $M_{\alpha}^2 = |M_{\alpha}^2| \exp(i \Psi)$ with $\Psi$ fixed and $|M_{\alpha}^2|$ are $(n+1)$ equidistant points covering the entire considered interval $0.65 \ {\rm GeV}^2 \leq |M^2| \leq 1.50 \ {\rm GeV}^2$. We took $n=85$, i.e., $|M_{\alpha}^2| = (0.65 + \alpha 0.01) \ {\rm GeV}^2$.}   
\bes
\label{condeno1}
\bea
\langle O_4 \rangle_{V+A} &=& (-0.00157  \pm 0.00030) \ {\rm GeV}^4
\nonumber\\
\Rightarrow \;
\langle a GG \rangle  &=& ( -0.0074 \pm 0.0018) \ {\rm GeV}^4 \ ,
\label{aGGo1}
\\
\langle O_{6} \rangle_{V+A} & = &  (+0.00136 \pm 0.00042) \ {\rm GeV}^6 \ .
\label{O6o1}
\eea
\ees
The central values were obtained by the method of minimal $\chi^2$ (with equal weights). The experimental uncertainties indicated above were obtained by an ``educated guess'' approach. Namely, the values of the condensates were varied around the obtained central values until the theoretical curve reached the outer edge of the experimental band. For example, the theoretical (dashed) curve in Fig.~\ref{FigPi64}(a) would reach the outer edges of the band for the first time when $\langle a GG \rangle  \approx ( -0.0074 \pm 0.0018) \ {\rm GeV}^4 $, the edges are reached at $|M^2| = 0.65 \ {\rm GeV}^2$. The theoretical (dashed) curve in Fig.~\ref{FigPi64}(b) would reach the upper outer edge for the first time when $\langle O_{6} \rangle_{V+A} =  (+0.00136 - 0.00042) \ {\rm GeV}^6$, the edge would be reached at $|M^2| = 0.65 \ {\rm GeV}^2$; for  $\langle O_{6} \rangle_{V+A}  =  (+0.00136 + 0.00042) \ {\rm GeV}^6$ the curve would reach the lower edge, at $|M^2| \approx 0.90 \ {\rm GeV}^2$.

In the approach with $\MSbar$ pQCD (+OPE), the corresponding obtained central values are $\langle a GG \rangle = +0.0050 \ {\rm GeV}^4$ and $\langle O_{6} \rangle_{V+A} = -0.00142 \ {\rm GeV}^6$. For simplicity, we will assume that the experimental uncertainties in the $\MSbar$ pQCD case are the same as in the $\A$QCD case mentioned above.

We can repeat all the sum rule analysis again for the cases of $\alpha_s(M_Z^2;\MSbar)=0.1185 \pm 0.0004$, i.e., the cases of the last two lines of Table \ref{tabres}. It turns out that the central values of the condensates change significantly when $\alpha_s(M_Z^2;\MSbar)$ is varied, but interestingly the quality of the fits (the values of $\chi^2$'s) does not change substantially. The final results for the condensates are
\bes
\label{conden}
\bea
\langle O_4 \rangle_{V+A} &=& -0.00157^{-0.00066}_{+0.00070} (\delta \alpha_s)  
\pm 0.00030({\rm exp}) \; [{\rm GeV}^4] 
\nonumber\\
\Rightarrow \;
\langle a GG \rangle  &=& -0.0074^{-0.0040}_{+0.0042}(\delta \alpha_s)  \pm 0.0018({\rm exp}) \; [{\rm GeV}^4]  = (-0.0074 \pm 0.0047) \ {\rm GeV}^4 \ ,
\label{aGG}
\\
\langle O_{6} \rangle_{V+A} & = &  +0.00136 \pm 0.00022(\delta \alpha_s) \pm 0.00042({\rm exp}) \; [{\rm GeV}^6] = (+0.00136 \pm 0.00047) \ {\rm GeV}^6 \ .
\label{O6}
\eea
\ees
When $\alpha_s(M_Z^2;\MSbar)$ decreases, $\langle a GG \rangle$ goes up and $\langle O_{6} \rangle_{V+A}$ goes down. For example, when $\alpha_s(M_Z^2;\MSbar)=0.1181$, the extracted central values are $\langle O_4 \rangle_{V+A} = -0.00087 \ {\rm GeV}^4$, $\langle a GG \rangle=-0.0032 \ {\rm GeV}^4$ and $\langle O_{6} \rangle_{V+A}=+0.00114 \ {\rm GeV}^6$. On the extreme right-hand side of Eqs.~(\ref{aGG})-(\ref{O6}) we added the variations in quadrature.

In the $\MSbar$ pQCD approach the results are
\bes
\label{condenMSbar}
\bea
\langle a GG \rangle(\MSbar)  &=& +0.0050^{-0.0020}_{+0.0019}(\delta \alpha_s)  \pm 0.0018({\rm exp}) \; [{\rm GeV}^4]  = (+0.0050 \pm 0.0027) \ {\rm GeV}^4 \ ,
\label{aGGMSbar}
\\
\langle O_{6} \rangle_{V+A}(\MSbar) & = &  -0.00142 \pm 0.00023(\delta \alpha_s) \pm 0.00042({\rm exp}) \; [{\rm GeV}^6] = (+0.00142 \pm 0.00048) \ {\rm GeV}^6 \ .
\label{O6MSbar}
\eea
\ees

We recall that the considered theory has as an input the value of the quantity $r_{\tau}(D=0)$, Eq.~(\ref{rtauD0cont}). The parameters of the $\A$ coupling were adjusted so that this value was $0.203$ by the method Eq.~(\ref{dan}) and $0.201$ by the method Eq.~(\ref{dBGan22}). A necessary check of consistency of our results would be to verify that these input values, and the extracted condensate value $\langle O_6 \rangle_{V+A} = +0.00136 \ {\rm GeV}^6$, are consistent with the OPAL data for $r_{\tau}$. This we can do in the following way. Applying the same type of sum rules, but now for the weight function $g(Q^2) = 2 (1 + Q^2/m_{\tau}^2)^2 (1 - 2  Q^2/m_{\tau}^2)$, leads to the following OPAL experimental $r_{\tau}$ value:
\bes
\label{rtauexp}
\bea
r_{\tau}(\Delta S=0; m_{u,d} =0)_{\rm exp} &=& 2 \int_{0}^{m_{\tau}^2} \frac{d \sigma}{m_{\tau}^2} \left( 1 - \frac{\sigma}{m_{\tau}^2} \right)^2 \left( 1 + 2  \frac{\sigma}{m_{\tau}^2} \right) \omega_{\rm exp}(\sigma) - 1  \approx 0.198 \pm 0.006 \ ,
\label{rtauexpa}
\\
\Rightarrow \;   r_{\tau}(D=0)_{\rm exp} &=& 2 \int_{0}^{m_{\tau}^2} \frac{d \sigma}{m_{\tau}^2} \left( 1 - \frac{\sigma}{m_{\tau}^2} \right)^2 \left( 1 + 2  \frac{\sigma}{m_{\tau}^2} \right) \omega_{\rm exp}(\sigma) - 1 + 12 \pi^2 \frac{\langle O_6 \rangle_{V+A}}{m_{\tau}^6}
  \nonumber\\
  & \approx & (0.198 \pm 0.006) + 0.005 = 0.203 \pm 0.006 \ ,
  \label{rtauexpb}
\eea
\ees  
where $(1.198 \pm 0.006)$ is the contribution of the integral over $\sigma$, and $0.005$ the contribution of the subtraction of the $D=6$ term.\footnote{There is no $D=4$ contribution to be subtracted, within our approximation of constant Wilson coefficients $(1 + {\cal C}_n a) \mapsto 1$. Further, OPAL data have the maximum available $\sigma$ equal to $\sigma_0=3.136 \ {\rm GeV}^2$ which is smaller than $m_{\tau}^2 =3.1572 \ {\rm GeV}^2$; this results in an error in the above integral of Eq.~(\ref{rtauexpa}) of $\delta r_{\tau} \sim 10^{-6}$, which is entirely negligible. The chirality-violating contributions $\Delta r_{\tau}(m_{u,d} \not=0)$ are not included in Eq.~(\ref{rtauexpa}), they involve an integral of $\omega^{(0)}(\sigma) = 2 \pi {\rm Im} \Pi^{(0)}_{V+A}(-\sigma - i \epsilon)$, and their leading part is $16 \pi^2 (m_u+m_d) \langle {\bar q} q \rangle/m_{\tau}^4 = - 8 \pi^2 f_{\pi}^2 m_{\pi}^2/m_{\tau}^4 \approx -0.003$, cf.~Refs.~\cite{Braaten2}.}
We recall that the theoretical central value for $r_{\tau}(D=0)$ was $0.203$ when using the method Eq.~(\ref{dan}) in Eq.~(\ref{rtauD0cont}), and it was $0.201$ when using the method Eq.~(\ref{dBGan22}) instead. We now see that this theoretical value is reproduced with very good precision again, now from OPAL data (and the subtraction of the higher-twist contribution), Eq.~(\ref{rtauexpb}). On the other hand, in $\MSbar$ pQCD approach, the experimental result is $r_{\tau}(D=0) = (0.198 \pm 0.006) - 0.005 = 0.193 \pm 0.006$ (because in this case the obtained $\langle O_6 \rangle_{V+A}$ was equal to $-0.00142 \ {\rm GeV}^6$). This differs significantly, by almost two standard deviations, from the $\MSbar$ pQCD theoretical central value for $ r_{\tau}(D=0)$, which was $0.182$ [cf.~discussion in the third paragraph after Eq.~(\ref{dBGan22})]. 

We point out that the reproduction of the value (\ref{rtauexpb}) by the considered $\A$-coupling theory ($\A$QCD) cannot be considered as a true prediction of the model ($\A$QCD+OPE), because this value was achieved by the correct adjustment of the parameters of the $\A$-coupling so that such value is in the end (consistently) reproduced. On the other hand, in $\MSbar$ pQCD + OPE, the coupling has no free parameters for adjustment (with the exception of ${\Lambda}_{\rm \MSbar}$ which is adjusted to give the correct world average at $Q^2=M_Z^2$), so $r_{\tau}(D=0)$ there is a prediction of the model, and this prediction is not very good.

\section{Predictions for some other physical quantities}
\label{sec:pred}

In the previous Section, we extracted from Borel sum rules for the $\tau$ semihadronic decays the values of the condensates $\langle a GG \rangle$ and $\langle O_{6} \rangle_{V+A}$, both for the considered $\A$-coupling framework ($\A$QCD) and for the usual pQCD $\MSbar$ approach.

Having now the coupling and some limited information about the higher-twist effects of the considered $\A$QCD theory, we may perform some additional checks. The theory is constructed in such a way as to agree with pQCD, in the considered scheme which agrees with the Lambert-MiniMOM scheme up to three-loop level. This scheme is not far away from the $\MSbar$ scheme; therefore, predictions of the $\A$QCD theory at high momenta $|Q^2| \gtrsim 10 \ {\rm GeV}^2$ are expected to be as good as in $\MSbar$ pQCD. However, as seen in the previous Section, the theory deviates from $\MSbar$ pQCD significantly in the $|Q^2| \sim 1 \ {\rm GeV}^2$ regime. It is in this regime that we can compare some predictions between the two theories.

The condensate values $\langle a GG \rangle$ and $\langle O_{6} \rangle_{V+A}$ were obtained in the previous Section from the Borel sum rules with $\Psi \equiv {\rm Arg} M^2 = \pi/6, \pi/4$, cf.~Figs.~\ref{FigPi64}. Therefore, strictly speaking, the results of the Borel sum rules with $\Psi = 0$, Fig.~\ref{FigPsi0} already represent predictions of the theory. As mentioned there, the resulting values $\chi^2$ for $\Psi=0$ are $1.3 \times 10^{-6}$ for $\A$QCD(+OPE) and $2.8 \times 10^{-5}$ for $\MSbar$ pQCD(+OPE).

\begin{figure}[htb] 
\begin{minipage}[b]{.49\linewidth}
  \centering\includegraphics[width=85mm]{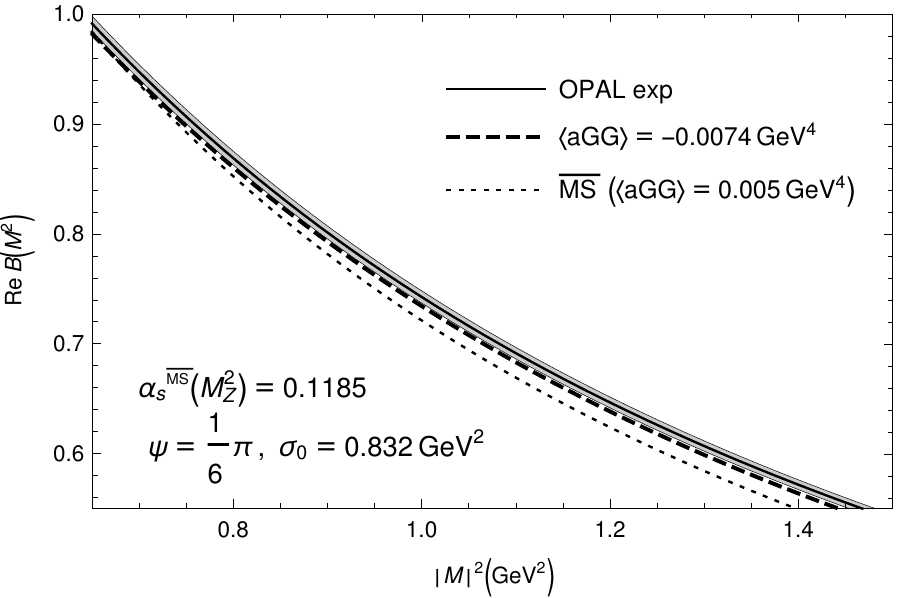}
  \end{minipage}
\begin{minipage}[b]{.49\linewidth}
  \centering\includegraphics[width=85mm]{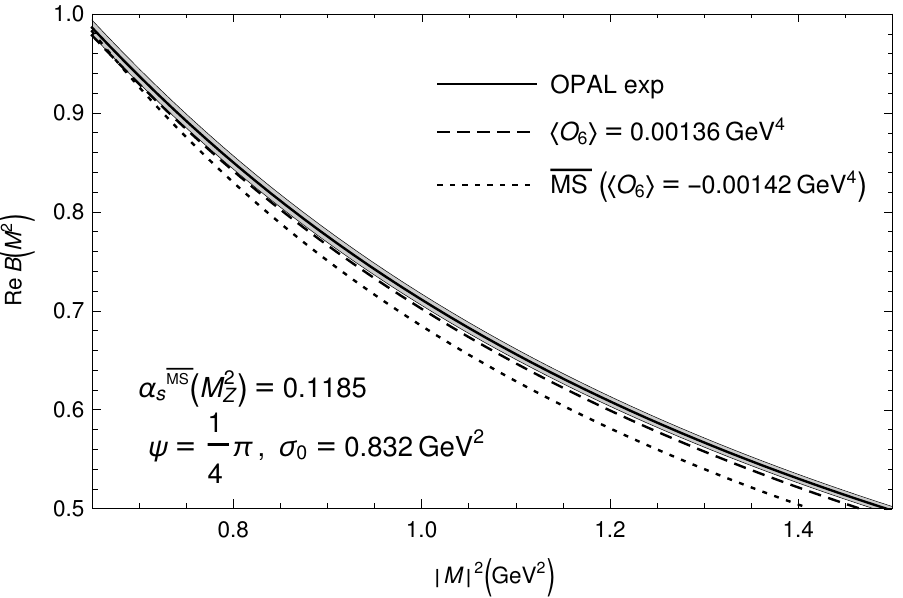}
\end{minipage}
\caption{\footnotesize The same as Figs.~\ref{FigPi64}, with the same values of condensates, but for the Borel sum rule upper bound $\sigma_0=0.832 \ {\rm GeV}^2$.}
\label{FiglowsPi64}
\end{figure}

\begin{figure}[htb] 
\centering\includegraphics[width=120mm]{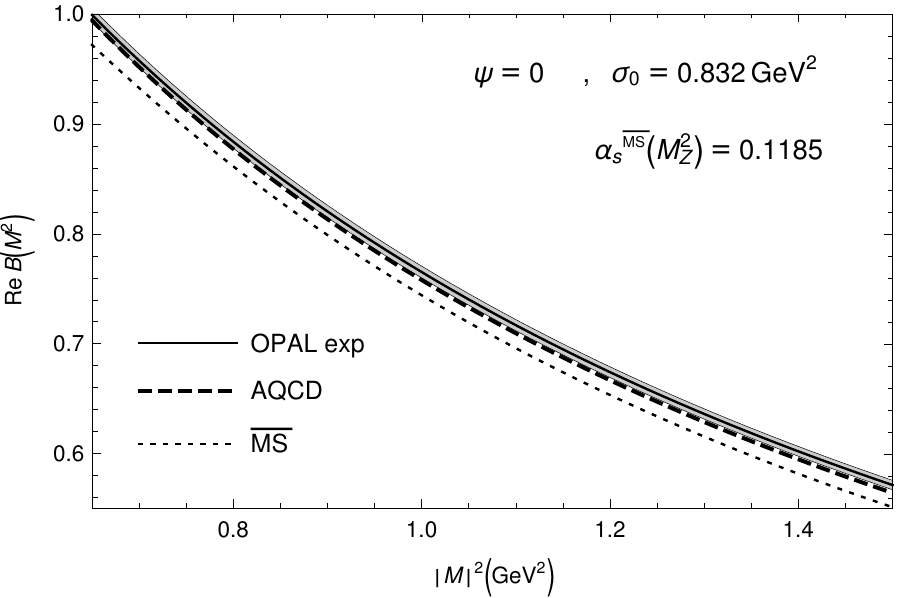}
\caption{\footnotesize The same as Fig.~\ref{FigPsi0}, with the same values of condensates, but for the Borel sum rule upper bound $\sigma_0=0.832 \ {\rm GeV}^2$.}
\label{FiglowsPsi0}
\end{figure}

Still within the Borel sum rules, we may want to check what happens with the quality of the reproduction of the Borel transforms when the upper integration bound $\sigma_0$ in the Borel sum rules changes, i.e., decreases. It turns out that if we decrease the upper integration bound $\sigma_0$ below $3.136 \ {\rm GeV}^2$, all the way down to $\sigma_0 \approx 1 \ {\rm GeV}^2$, and keep the same condensate values as obtained in the $\sigma_0 =3.136 \ {\rm GeV}^2$ case, the quality of the curves in comparison with the experimental curves does not deteriorate significantly when $\A(Q^2)$ +OPE (AQCD) is used, but it does deteriorate significantly when $\MSbar$ pQCD+OPE is used \cite{inprep}. In Figs.~\ref{FiglowsPi64}, \ref{FiglowsPsi0} we present these results for the very low value $\sigma_0=0.832 \ {\rm GeV}^2$. If decreasing $\sigma_0$ even further, the quality of the $\A$QCD fit deteriorates, too; this is to be expected, because for such low values the spectral function is dominated by the $\rho$ resonance, cf.~Fig.~\ref{FigOPAL}.

We now present yet another prediction of the theory, this time for the (properly normalized) production ratio $R(\sigma)$ for $e^+ e^- \to$ hadrons, which is a timelike observable. Instead of considering the experimental values of $R(\sigma)$, it is more convenient to consider the related experimental values of the vector (V) channel Adler function ${\cal D}_V(Q^2)$, which is a spacelike observable and can thus be evaluated as described in the previous Sections. The experimental values are obtained by the integral transformation
\be
{\cal D}_{V, {\rm exp}}(Q^2) = Q^2 \int_{4 m_{\pi}^2}^{\sigma_0} \frac{R_{\rm data}(\sigma)}{(\sigma + Q^2)^2} d \sigma + Q^2 \int_{\sigma_0}^{\infty} \frac{R_{\rm pert}(\sigma)}{(\sigma + Q^2)^2} d \sigma,
\label{DV}
\ee
where $\sigma_0$ is a sufficiently high (squared) energy scale above which the perturbative evaluation of $R(\sigma)$ is good. This experimental function ${\cal D}_{V, {\rm exp}}(Q^2)$ was obtained from the experimental data $R_{\rm data}(\sigma)$ in Refs.~\cite{Eidel,Nest3a}, cf.~also Ref.~\cite{NestBook}. 
Here,  ${\cal D}_V(Q^2) = 1 + d_V(Q^2)$ where the dimension $D=0$ (leading-twist and massless) part $d_V(Q^2;D=0) \equiv d(Q^2,D=0)$ is the same as in the V+A case, Eqs.~(\ref{dpt})-(\ref{dBGan22}) and Eq.~(\ref{Adlfull}). The OPE expansion has the form of Eq.~(\ref{Adlfull}), where now the condensates are for the V channel only and have an additional overall factor 2 due to the normalization convention
\bea
{\cal D}_V(Q^2) &\equiv&  - 4 \pi^2 \frac{d \Pi_V(Q^2)}{d \ln Q^2} 
 =  1 + d(Q^2;D=0) + 2 \pi^2 \sum_{n \geq 2}
 \frac{ n 2 \langle O_{2n} \rangle_V}{(Q^2)^n}  \ ,
\label{AdlVfull}
\eea
where the convention (notation) $\langle O_{2n} \rangle_{V+A} = \langle O_{2n} \rangle_{V} + \langle O_{2n} \rangle_{A}$ is maintained. In the approximations as applied in Eqs.~(\ref{aGGO4}), we have \cite{Braaten2}
\be
2 \langle O_4 \rangle_V =  2 \langle O_4 \rangle_A  = \langle O_4 \rangle_{V+A} \ .
\label{O4V}
\ee
This means that $2 \langle O_4 \rangle_V = -0.0074 \ {\rm GeV}^4$ in the $\A$QCD case, and $0.0050 \ {\rm GeV}^4$ in $\MSbar$ pQCD case, using the extracted values of $\langle O_4 \rangle_{V+A}$ in the previous Section in the central case $\alpha_s(M_Z^2;\MSbar)=0.1185$, Eqs.~(\ref{aGG}) and (\ref{aGGMSbar}). In order to extract the value of $\langle O_6 \rangle_V$ from our obtained values $\langle O_6 \rangle_{V+A}$, we can use the factorization hypothesis \cite{Ioffe2} which gives
\bes
\label{O6V}
\bea
\langle O_6 \rangle_{V+A} & \approx & +\frac{128}{81} \pi^2 a(\mu^2) \langle {\bar q} q \rangle_{\mu}^2, \quad
\langle O_6 \rangle_{V-A} \approx  -\frac{64}{9} \pi^2 a(\mu^2) \langle {\bar q} q \rangle_{\mu}^2,
\label{O6Va}
\\
\Rightarrow \;
\langle O_6 \rangle_V &\approx&  -\frac{224}{81} \pi^2 a(\mu^2) \langle {\bar q} q \rangle_{\mu}^2 \approx - \frac{7}{4}  \langle O_6 \rangle_{V+A}.
\label{O6Vb}
\eea
\ees
The factorization hypothesis has relative error $\approx 1/N_c^2 \approx 10 \%$. Using the relation (\ref{O6Vb}) and the central extracted values of $\langle O_6 \rangle_{V+A}$ of the previous Section, Eqs.~(\ref{O6}) and (\ref{O6MSbar}), we obtain $\langle O_6 \rangle_V \approx -0.0024 \ {\rm GeV}^6$ in the $\A$QCD case, and $\approx +0.0025$ in the $\MSbar$ pQCD case. Using these values in Eq.~(\ref{AdlVfull}), we obtain the theoretical predictions for ${\cal D}_V(Q^2)$. 

\begin{figure}[htb] 
  \centering\includegraphics[width=120mm]{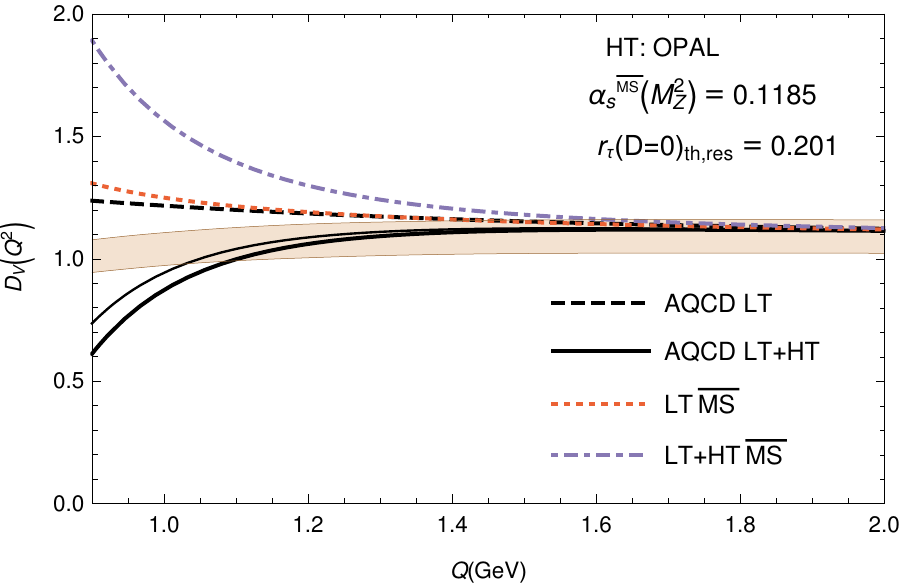}
  \caption{\footnotesize The V-channel Adler function at low positive values of $Q^2$ (where $Q \equiv \sqrt{Q^2}$): the grey band (online: brown band) represents the experimental values. The theoretical curves are for $\alpha_s(M_Z^2)=0.1185$. The thick black line represents $\A$QCD results with $D=4$ and $D=6$ terms (higher-twist) included; the dashed line is only the leading-twist result. Included are also the corresponding lines for the $\MSbar$(+OPE) approach, with the corresponding $\MSbar$ condensates. For comparison, the $\A$QCD result  for the lower value $\alpha_s(M_Z^2;\MSbar)=0.1181$ and with the corresponding higher-twist terms is included as the narrower black line. $N_f=3$ is used throughout. The experimental band was taken from Ref.~\cite{NestBook} (Fig. 1.7 there), cf.~also Refs.~\cite{Nest3a}.}
  \label{figAdler}
\end{figure}

In Fig.~\ref{figAdler} we present the corresponding experimental results and the theoretical predictions in the considered $\A$QCD+OPE and $\MSbar$ pQCD+OPE, in an intermediate IR regime of $0.9 \ {\rm GeV}^2 < Q^2 < 2.0 \ {\rm GeV}^2$. The higher-twist (HT) terms are the mentioned $D=4$ and $D=6$ terms. In $\A$QCD, the term $d(Q^2,D=0)$ was evaluated with the resummation method Eq.~(\ref{dBGan22}), although the TPS method Eq.~(\ref{dan}) gives very similar results.
In the Figure we can see that the present $\A$QCD+OPE approach gives results compatible with $D_{V {\rm exp}}(Q^2)$ for scales down to $Q^2 \approx 1 \ {\rm GeV}^2$; at $Q^2 < 1 \ {\rm GeV}^2$ this approach is expected to fail, because the OPE terms $\sim 1/(Q^2)^n$ cannot describe such deep IR regime. On the other hand, the $\MSbar$ pQCD + OPE approach appears to fail already at the scales $Q^2 \approx 2.6 \ {\rm GeV}^2$ ($Q \approx 1.6$ GeV). Moreover, inclusion of the higher-twist terms ($D=4$ and $D=6$) clearly improves the results in the $\A$QCD+OPE approach, while in the $\MSbar$ pQCD + OPE approach the corresponding higher-twist terms deteriorate the results. Numerically, this has to do with the fact that the signs of the extracted $D=4, 6$ condensates in the $\MSbar$ pQCD+OPE approach are opposite to those of the $\A$QCD+OPE approach, cf.~the previous Section. We also see that for the lower value $\alpha_s(M_Z^2;\MSbar)=0.1181$, the results of $\A$QCD with the corresponding higher-twist terms [$\langle O_4 \rangle = -0.00087 \ {\rm GeV}^4$ and  $\langle O_6 \rangle= +0.00114 \ {\rm GeV}^6$, Eqs.~(\ref{conden})] agree even better with the experimental band at low $Q^2$ (the narrower black line).

We wish to stress that the values of the gluon condensate  $\langle a GG \rangle$, quoted in the literature and obtained from various sum rule applications (in $\MSbar$), vary strongly, from positive to negative values. In the original work on the sum rules \cite{Shifman:1978bx}, a positive value $0.012  \ {\rm GeV}^4$ was obtained, from charmonium physics.
\begin{table}
\caption{The gluon condensate values from the literature.}
\label{tabgl}  
\begin{ruledtabular}
\begin{tabular}{r l}
  $\langle a GG \rangle$ $[{\rm GeV}^4]$ &  work and method
  \\
  \hline
  $0.012$ & \cite{Shifman:1978bx},  charmonimum sum rules
  \\
  $0.037 \pm 0.015$ & \cite{Doming14}, Finite Energy Sum Rules in $e^+e^-$
  \\
  $0.022 \pm 0.004$ & \cite{O4sr1}, QCD-moment sum rules
  \\
  $0.024 \pm 0.006$ & \cite{O4sr1}, QCD-exponential moment sum rules
  \\
  $0.007 \pm 0.005$ & \cite{anOPE}, $\tau$ V+A sum rules with holomorphic coupling, $\A(0)>0$
  \\
  $0.012 \pm 0.005$ & \cite{anpQCD2}, holomorphic pQCD (not $\MSbar$) coupling with $a(0)>0$
  \\
  $0.077 \pm 0.087$ & \cite{BaBaPi1}, stochastic pQCD approach for $SU(3)$ plaquette
  \\
  $0.006 \pm 0.012$ & \cite{Ioffe1}, three-loop Borel sum rules
    \\
  $0.005 \pm 0.004$ &  \cite{Ioffe2}, charmonium sum rules
  \\
  $-0.005 \pm 0.003$ &  \cite{ALEPHupdate}, V-channel $\tau$ decay multiparameter fit
  \\
  $-0.034 \pm 0.004$ & \cite{ALEPHupdate}, A-channel $\tau$ decay multiparameter fit
  \\
  $-0.020 \pm 0.003$ & \cite{ALEPHupdate}, V+A-channel $\tau$ decay multiparameter fit
  \\
  $-0.007 \pm 0.005$ & this work, $\tau$ V+A sum rules with holomorphic coupling, $\A(0)=0$
\end{tabular}
\end{ruledtabular}
\end{table}
Finite energy sum rules (FESR) give usually positive values. For example, FESR for $e^+e^-$ annihilation in the $c$-quark region gave $0.037 \pm 0.015 \ {\rm GeV}^4$ \cite{Doming14}. QCD-moment and QCD-exponential moment sum rules for heavy quarkonia also gave positive values, $\langle a G G \rangle \approx (0.022 \pm 0.004) \ {\rm GeV}^4$ and $(0.024 \pm 0.006) \ {\rm GeV}^4$, respectively, Refs.~\cite{O4sr1}. A model with holomorphic coupling $\A(Q^2)$ and finite positive value $\A(0)$ \cite{2danQCD} also gave positive values $\langle a GG \rangle=(0.007 \pm 0.005) \ {\rm GeV}^4$ \cite{anOPE}, and a pQCD holomorphic coupling $a(Q^2)$ \cite{anpQCD2} with finite positive $a(0)$ (not in $\MSbar$) gave similar values $(0.012 \pm 0.005) \ {\rm GeV}^4$.\footnote{
Here we added the spontaneous chiral symmetry breaking correction $6 f_{\pi}^2 m_{\pi}^2 \approx 0.002 \ {\rm GeV}^4$ [cf.~Eq.~(\ref{aGGO4b})] to the gluon condensate values obtained in Refs.~\cite{anOPE,anpQCD2}.}
  Calculation of the plaquette in $SU(3)$ to high orders \cite{BaBaPi1}, in numerical stochastic pQCD approach, gives $\langle a GG \rangle = 0.077 \pm 0.087 \ {\rm GeV}^4$ \cite{BaBaPi2} (this calculation was performed to sufficiently high order to see the onset of the dominance of the $D=4$ renormalon in pQCD).
On the other hand, in Ref.~\cite{Ioffe1} three-loop Borel sum rules (in $\MSbar$) for $\tau$ decay data in $V+A$ channel gave values $(0.006 \pm 0.012) \ {\rm GeV}^4$, compatible with zero. In Ref.~\cite{Ioffe2}, where charmonium sum rules were included in the analysis, similar values  $(0.005 \pm 0.004) \ {\rm GeV}^4$ were obtained. However, an (updated) multiparameter fit of ALEPH $\tau$ decay data gave consistently negative values for $\langle a G G \rangle$: $(-0.005 \pm 0.003) \ {\rm GeV}^4$ for $V$ channel,  $(-0.034 \pm 0.004) \ {\rm GeV}^4$ for $A$ channel, and $(-0.020 \pm 0.003) \ {\rm GeV}^4$ for $V+A$ channel \cite{ALEPHupdate} (Table 4 there). We summarize  the comparison of the mentioned gluon condensate values in Table \ref{tabgl}.

In view of the fact that there exist various interpretations of gluon and quark condensates, it is not clear whether the gluon condensate should be positive or negative. For example, the authors of Ref.~\cite{BrodShr} argue that the QCD condensates are not associated with the vacuum, but with the internal dynamics of hadrons and color confinement, and thus give zero contribution to the cosmological constant. Another general aspect of the fitted values of condensates is that they may change significantly when the number of terms taken in the low-energy leading-twist contribution increases \cite{KatParSid}.
Further, we point out that the values of the gluon condensate in the literature, mentioned in the previous paragraph, were obtained with QCD couplings with either singular behavior in the infrared or with finite positive value $\A(0)$, while the QCD coupling considered here has $\A(0)=0$ and thus does not have a well defined beta function $\beta(\A)$.

\section{Summary}
\label{sec:summ}

In this work we constructed a QCD running coupling $\A(Q^2)$ which fulfills a host of high- and low-energy restrictions. It was constructed in the $N_f=3$ Lambert-MiniMOM renormalization scheme, which coincides with the $N_f=3$ lattice MiniMOM renormalization scheme (MM) \cite{MiniMOM} at three-loop level. The usual scaling was used ($\Lambda_{\rm \MSbar}$ instead of $\Lambda_{\rm MM}$). The coupling $\A(Q^2)$ was constructed in such a way that it automatically reflects the holomorphic (analytic) behavior of the spacelike QCD observables ${\cal D}(Q^2)$ in the complex $Q^2$-plane. At high momenta $|Q^2| > 4 \ {\rm GeV}^2$ the coupling $\A(Q^2)$ was required to practically merge with its pQCD analog $a(Q^2)$ [$\equiv \alpha_s(Q^2)/\pi$] in the same scheme, while at very low momenta $0 \leq Q^2 \alt 1 \ {\rm GeV}^2$ it was required to reproduce qualitatively the main features suggested by the lattice calculations. At the low momentum scales $|Q^2| \sim m_{\tau}^2$, the obtained $\A$QCD theory was required to reproduce the correct measured value of the semihadronic strangeless and massles $\tau$ decay ratio $r_{\tau} \approx 0.20$ [where $r_{\tau}$ is canonical: $r_{\tau} = a + {\cal O}(a^2)$]. Application of the $\A$QCD+OPE approach to the Borel sum rules for the $V+A$ spectral functions of $\tau$-lepton, with upper bound $\sigma_0 \approx m_{\tau}^2$ and in specific directions (rays) in the complex plane of the Borel scale $M^2$, then yields specific values, both for the gluon condensate $\langle a GG \rangle$ ($\approx - 0.007 \pm 0.005 \ {\rm GeV}^4$) and for the condensate $\langle O_{6} \rangle_{V+A}$ ($\approx  + 0.0014 \pm 0.0005 \ {\rm GeV}^6$). Subsequently, with these condensate values, very good agreement with the measured values is obtained for the Borel sum rules $B(M^2)$ along the positive axis of $M^2$, significantly better than in the usual pQCD $\MSbar$+OPE approach. Verification of the Borel sum rules with significantly lower upper bound, $\sigma_0 \approx 0.83 \ {\rm GeV}^2$, still gives good agreement between the experimental and theoretical predictions, in contrast to the pQCD $\MSbar$+OPE approach. Furthermore, application of the obtained theoretical results to the $V$-channel Adler function, closely related to the production ratio $R(\sigma)$ of the $e^+e^- \to$ hadrons, yields predictions significantly closer to the experimental results than the pQCD $\MSbar$+OPE approach. The Mathematica scripts which calculate the considered coupling $\A(Q^2)$ and its higher-order analogs are freely available \cite{4danQCDcoupl}.

\begin{acknowledgments}
G.C.~acknowledges the support by FONDECYT (Chile) Grant No.~1130599; C.A.~acknowledges the support by the Spanish Government and ERDF funds from the EU Commission [Grant No.~FPA2014-53631-C2-1-P] and by CONICYT Fellowship ``Becas Chile'' Grant No.~74150052. We are grateful to S.~Peris and A.~Sternbeck for valuable discussions and clarifications.
\end{acknowledgments}

\end{document}